\documentclass[
nofootinbib,
 amsmath,amssymb,
 aps,
superscriptaddress]{revtex4-2}

\usepackage{graphicx}
\usepackage{dcolumn}
\usepackage{bm}
\usepackage{soul}

\usepackage{color} 
\usepackage{hyperref}
\hypersetup{
    colorlinks=true, 
    pdfborder = {0 0 0.5 [3 3]},
    anchorcolor=black,
    citecolor=blue,
    linktoc=all,    
    linktocpage=true,
    linkcolor=red,
	urlcolor=blue
}

\usepackage{tikz}
\usetikzlibrary{mindmap,shadows,positioning}
\definecolor{myTeal}{HTML}{A4C8C8}

\begin{document}

\title{Solar system tests and neutron stars in $f(R)$ gravity revisited}




\author{Hodek M. García}
\email{hodek.mealstrom@ciencias.unam.mx}
\address{Instituto de Ciencias Nucleares, Universidad Nacional Autónoma de México, Circuito Exterior CU, A.P. 70-543, CDMX 04510, México}
 \author{Marcelo Salgado}%
 \email{marcelo@nucleares.unam.mx}
\address{Instituto de Ciencias Nucleares, Universidad Nacional Autónoma de México, A.P. 70543, México CDMX. 04510, México} 
\address{Departament  de  F\'isica,  Universitat  de  les  Illes  Balears,  Palma  de  Mallorca,  E-07122,  Spain}




\date{\today}

\begin{abstract}
By implementing a full non-linear treatment of $f(R)$ 
gravity in static and spherically symmetric spacetimes, 
we analyze two scenarios. The first one within the context of the solar-system tests where 
we try to recover the chameleon effects without any approximations in the equations (e.g. linearization) 
from $f(R)$ models that are compatible with cosmology. The second scenario deals with a quadratic $f(R)$ model that is tested in neutron stars. 
This scenario, which is associated with strong gravity, is completely independent from the first one, but exploits the fact that the equations and formalism are basically the same in both applications. The difference between the two goals lies mainly in the values of the constants involved in the specific $f(R)$ models and the equation of state (EOS) of the central object (Sun or neutron star), but the numerical techniques and the general form of the field equations remain valid in both situations. For the neutron star problem we employ for the first time and in the context of $f(R)$ gravity a multiple algebraic polytropic EOS that mimics accurately realistic EOS in several density ranges. By doing so we avoid the numerical interpolation needed when a realistic EOS is given in tabulated form. Furthermore, we compare our results with the latest data, which includes the  most massive neutron star known to date of about $2.35 M_\odot$ from PSRJ0952-0607.
\end{abstract}



\maketitle


\section{\label{sec:level1} Introduction }
Since the discovery of the accelerated expansion of the universe by several independent observations  \cite{riessObservationalEvidenceSupernovae1998,perlmutterMeasurementsOmegaLambda1999, amanullahSpectraLightCurves2010,weinbergObservationalProbesCosmic2013} a plethora of studies in cosmology have been devoted  to understand and explain this late-time acceleration era.\footnote{Not to mention the other phase of cosmic acceleration which is believed to have occurred prior to the radiation domination era, i.e., during the stages of evolution of the universe also known as \textit{inflation} (see \cite{liddleCosmologicalInflationLargeScale2000} and references therein). } The current paradigm of the large-scale structure of the universe that explains this phenomenon appeals to the existence of a cosmological constant  $\Lambda$, better known as \textit{dark energy} (DE),
which produces a negative pressure and therefore a positive acceleration 
as the universe expands. Prior to this era, the universe was matter dominated, 
where another \textit{dark matter} (DM) component seems necessary. Both components account for the dark sector of the universe which has been thoroughly modeled by many proposals \cite{copelandDynamicsDarkEnergy2006,silvestriApproachesUnderstandingCosmic2009,arbeyDM2021} 
and has been measured indirectly with several independent gravitational observations. In particular, if General Relativity (GR), which is  widely accepted as the fundamental (classical) theory of gravity \cite{waldGeneralRelativity1984}, is assumed, together with 
the inflationary prediction $\Omega=1$ (total energy-density of the universe in units of critical energy-density) an intriguing picture of the universe emerges. The dark sector accounts for  approximately 96$\%$ of its total material content, with $\sim 70\%$ given by the dark energy fluid of negative pressure,  whose origin is currently a mystery, and the rest $\sim 26\%$ by (cold) dark matter (CDM). Only $\sim 4\%$ of the total budget  $\Omega$ is associated with ordinary (baryon) matter.
The resulting description has become known as the concordance $\Lambda$CDM model  \cite{bahcallCosmicTriangleRevealing1999}. Although this 
cosmological picture matches  the current observations of the universe, it is plagued  by several theoretical and epistemological issues including the well known fine-tuning problem related to the vacuum energy scale \cite{carrollCosmologicalConstant2001}, 
and the unknown nature of DM, whose effects are most notably seen in the dynamics of (spiral) galaxies, in the clustering of large-structure and also as imprints in the CMB. Direct searches for DM have failed so far, to the extent that the scientific community has proposed instead modified models of gravity, such as MOND to explain the galactic dynamics without the need of DM \cite{MOND}.

As mentioned above, the cosmological constant is so far the simplest model to explain the late accelerated expansion of the universe, even if a deeper theoretical explanation about its origin is missing. Another prediction of this DE model is its {\it equation of state} (EOS) $\omega_{DE}=-1$ throughout the history of the universe. Current astronomical experiments (e.g. DESI \cite{DESI2024,DESI2025}) are trying to probe the nature of this dark energy. For instance, if $\omega_{DE}\neq - 1$ at some epoch, then such data would point to a DE component different from $\Lambda$. In this regard, 
several theoretical attempts have been put forward to describe the DE component since the SNIa distance measurements \cite{perlmutterMeasurementsOmegaLambda1999,riessObservationalEvidenceSupernovae1998}. 
Some of the most analyzed ones in this direction are {\it quintessence} and {\it k-essence}, which are scalar-field models with different kinds of potentials and kinetic terms, respectively 
\cite{copelandDynamicsDarkEnergy2006,tsujikawaquint2013}. More recently, a geometric attempt 
to generate dynamically a late cosmological constant is $f(R)$ gravity, where a non-linear Lagrangian of the Ricci scalar is proposed  (see \cite{nojiriCanGravityBe2008,nojiriIntroductionModifiedGravity2006,straumannProblemsModifiedTheories2008,capozzielloBirdEyeView2010,nojiriModifiedGravityTheories2011pr,cliftonModifiedGravityCosmology2012,sotiriouFRTheoriesGravity2010,defeliceFRTheories2010,nojiriModifiedGravityTheories2017,jaimeCosmologyRevisited2012} for a review). In particular, a prediction of $f(R)$ gravity is the EOS for (geometric) DE which changes in cosmic time \cite{jaimeCosmologyRevisited2012,jaimeNoteEquationState2014}. The $\Lambda {\rm CDM}$ model is represented by $f(R)= R-2\Lambda$ together with the ordinary (baryon) and DM sectors. In this respect, it is worth stressing that the quadratic model $f(R)= R+\alpha R^2$ 
(with $\alpha$ a suitable constant with square length units) was proposed by Starobinsky to explain geometrically the inflationary period \cite{starobinskyNewTypeIsotropic1980}. 
However, like several other modified theories of gravity, $f(R)$ gravity has the problem that there is no fundamental principle that singles out this function. Clearly, simplicity favors $\Lambda {\rm CDM}$, which is associated with GR. But apart from that, other $f(R)$ proposals (not motivated by quantum-gravity corrections) have been constructed by ``engineering'' and trial and error. Moreover, there is the following conceptual misunderstanding that has led to a long debate in the past. When $f(R)$ models are presented as a scalar-tensor theory, they predict the absence of a kinetic term for the scalar-field, which amounts to a Brans-Dicke parameter $\omega_{BD}\equiv 0$ 
\cite{chibaGravityScalarTensorGravity2003,chibaSolarSystemConstraints2007}. However, we know from observations in the solar system that $\omega_{BD}\approx 4\times 10^4$ in order for a post-Newtonian parameter $\gamma\approx 1$ \cite{bertottiTestGeneralRelativity2003}. A way out of this conundrum is that in reality $f(R)$ models, when viewed as scalar-tensor theories, 
are endowed with a scalar-field potential that in some circumstances can suppress the scalar degree of freedom in the neighborhood of the solar system so that $\gamma\approx 1$. This is the so-called {\it chameleon} mechanism. In view of this, not all $f(R)$ models are ruled out automatically, but only those 
that do not develop the chameleon effect. Originally, this mechanism was proposed in a different context in connection with potential violations of the Einstein's {\it equivalence principle} \cite{khouryChameleonCosmology2004}, and later ``exported'' to the $f(R)$ framework. The chameleon mechanism is an inherent non-linear effect that involves different energy or length scales. Thus, even in the simplest scenarios it is difficult to 
generate, and require numerical methods  or heuristic simplifications that entails modeling the actual scalar-field potential by gluing different potentials in regions occupied by ordinary matter (like in the Earth, Sun, Moon, and tabletop experiments) \cite{kraiselburdThickShellRegime2019,kraiselburdEquivalencePrincipleChameleon2018,motaEvadingEquivalencePrinciple2007}. So far, only a few successful $f(R)$ cosmological models are able to develop the chameleon mechanism in the solar system
\cite{negrelliSolarSystemTests2020,huModelsCosmicAcceleration2007}. As stressed above, the scales involved (cosmological and local) make its mathematical and numerical treatment a real challenge. Only by using semi-analytic and semi-linear treatments has it been possible to recover the chameleon, {\it thin-shell} or {\it screening}, as this effect is also known.

The goal of this paper is twofold: the first one is to analyze  $f(R)$ gravity in a full non-linear fashion in the solar system (assuming spherical symmetry) and show if the chameleon mechanism does or does not emerge naturally from several specific $f(R)$ models. Unfortunately, we will see that even with an {\it arbitrary-precision} numerical code, it was impossible just to reproduce the actual conditions of the Sun with the $f(R)$ models having a cosmological built-in scale and without approximations in the equations. Therefore we were unable to check if the chameleon effect really appears or not, as indeed happens when one uses linear or semilinear treatments \cite{negrelliSolarSystemTests2020,huModelsCosmicAcceleration2007}. Clearly, this {\it negative} or {\it null result} does not prove that one cannot succeed in reproducing the solar system tests in a full non-linear fashion using other algorithms. But the problem remains open.

The second goal consists in analyzing the consequences of a different kind of $f(R)$ models, a quadratic model (which does not have a cosmic built-in scale) in neutron stars, regardless if such models are able or not to produce a late cosmic acceleration. In this case, the parameter involved in the $f(R)$ theory is of the same order of magnitude of the scale associated with neutron stars. Therefore, when solving the full non-linear equations (also in the spherically symmetric approximation) the problem of a density contrast between the value of the parameter and the density of a neutron star is not present. Thus, the numerical treatment becomes less challenging than in the solar system scenario and becomes more manageable. We will show that the quadratic-gravity model can produce neutron stars with masses larger than in GR when using the same EOS for the nuclear matter (polytropes that accurately mimic realistic EOS). This opens the possibility that this kind of gravity model can account for observed neutron stars with large masses (e.g. as large as $\sim 2.35 M_\odot$ \cite{romani2022}) without the need for ``exotic'' EOS. Needless to say, such gravity model would also need to face the rest of gravitational tests of GR if one is to take it seriously as a universal gravity theory. 

The paper is organized as follows: Sec. \ref{sec:framework} presents the general formalism and the explicit differential equations in $f(R)$ gravity describing static and spherically symmetric spacetimes. It also introduces the specific  $f(R)$ models to be tested in the solar system scenario. Sec. \ref{sec:solarsys} discusses the numerical results from the integration of the equations 
for the Sun. Sec. \ref{sec:starsR2} deals with the analysis of neutron stars in the quadratic-gravity model. In particular, we describe the three EOSs used to 
represent the nuclear matter within those objects, and show the numerical results from integrating the equations of hydrostatic equilibrium, including the total mass of the star with respect to the central density and with respect the radius (i.e. the mass-density and mass-radius curves), as well as the solutions for 
the Ricci scalar and the metric functions with respect to the radial coordinate $r$. We also include the 
analysis of NS using an {\it incompressible fluid.} Finally, Sec. \ref{sec: conclusion} presents our conclusions and final remarks. Supplemental material that helps to 
clarify some aspects of the main text can be found in five Appendices.



\section{Framework and formalism of metric $f(R)$ gravity}
\label{sec:framework}
In what follows, we briefly describe the main features of $f(R)$ theory under the metric approach as opposed to the Palatini formalism \cite{sotiriouFRTheoriesGravity2010}. For GR both approaches are equivalent, but it is not the case for $f(R)$ gravity, in general. Moreover, we also follow the method developed in \cite{jaimeRobustApproachGravity2011},  which circumvents the (conformal) transformation of $f(R)$ theory into a 
scalar-tensor theory (STT) in the Einstein frame (see \cite{canateSphericallySymmetricBlack2016}).

 The general action for an $f(R)$ theory of gravity in a 4-dimensional manifold   is given by 
\begin{equation}
\label{eq::fR_formalism_action_1}
 S[g_{ab}, \Psi] = \int \frac{f(R)}{2\kappa} \sqrt{-g} d ^4x+ S_{M}[g_{ab},\Psi],
\end{equation}
where $\kappa = 8\pi G/c^4$. 
Varying  the action with respect to the metric gives 
\begin{equation}
\label{eq::fR_formalism_1}
    f_R R_{ab}-\frac{1}{2}f g_{ab}-(\nabla_a\nabla_b-g_{ab}\Box )f_R =\kappa T_{ab},
\end{equation}
where $f_R := \partial_R f$, $\Box := g^{ab}\nabla_a\nabla_b$, and $T_{ab}$ is the energy-momentum tensor (EMT) of matter that is represented schematically by $\Psi$ in the action functional. In our specific applications that we analyze below, we will consider a perfect fluid for the matter content.
Moreover, the trace of Eq. \eqref{eq::fR_formalism_action_1} yields
\begin{equation}
     \label{eq::fR_formalism_3}
     \Box R = \frac{1}{3f_{RR}}\left[\kappa T -3 f_{RRR} (\nabla R)^2 + 2f - Rf_R\right],
\end{equation}
where $T:= T^a_{\,\,a}$. Thus, we have explicitly obtained a scalar equation for $R$. Furthermore, using Eq. \eqref{eq::fR_formalism_3} in Eq. \eqref{eq::fR_formalism_1} we obtain 
\begin{equation}
\label{eq::fR_formalism_4}
    G_{ab} = \frac{1}{f_R}\left[f_{RR}\nabla_a\nabla_b  R + f_{RRR}(\nabla_a R)(\nabla_b R) - \frac{g_{ab}}{6} \left(Rf_R + f  + 2\kappa T\right) + \kappa T_{ab}\right],
\end{equation}
where $G_{ab} = R_{ab} - g_{ab}R/2$ is the Einstein tensor, $(\nabla R)^2:= g^{ab}(\nabla_a R)(\nabla_b R)$, $f_{RR}:=\partial^2_R f$ and $f_{RRR}:=\partial^3_{R} f$. Eqs. \eqref{eq::fR_formalism_4} and  \eqref{eq::fR_formalism_3}  are the fundamental coupled field equations of $f(R)$ gravity. 
 In this way, the theory can be described in terms of two second order partial differential equations (PDE), one for the metric \eqref{eq::fR_formalism_4} and another one for the Ricci scalar \eqref{eq::fR_formalism_3}, instead of one fourth order tensor PDE for the metric alone.  Under this approach the theory has an extra scalar degree of freedom represented by the Ricci scalar itself. Unlike GR, where $R$ and $T$ are algebraically related by $R=-\kappa T$, in $f(R)$ gravity this relationship is a differential one given by Eq. \eqref{eq::fR_formalism_3}. The only exception to this remark is precisely when $f(R) = R- 2\Lambda$, which corresponds to GR with $\Lambda$. Written in second order form, we can treat the theory in a more cleaner and straightforward way relative to the fourth order method or even relative to the scalar-tensor approach, where the scalar degree of freedom $\phi= f_R$ is defined and then one is required to invert everywhere (when it is possible) $R$ in terms of $\phi$. Furthermore, we avoid also the conformal transformation to the so called {\it Einstein frame}, which uses another scalar field related with 
 $\phi$ and a conformal metric. The conformal STT approach is, however, often used in $f(R)$ gravity, in particular, when studying compact objects, like neutron stars \cite{yazadjievNonperturbativeSelfconsistentModels2014, kobayashiCanHigherCurvature2009, babichevRelativisticStarsGravity2009, upadhyeExistenceRelativisticStars2009}.

It is instrumental to define the following ``potential" and its derivative from Eq. \eqref{eq::fR_formalism_3}: 
 \begin{equation}
 \label{dVeff}
    \frac{d V^{\text{eff}}(R,T)}{dR} =\frac{\kappa T + 2f - Rf_R}{3},
\end{equation}
where we assume that $f_{RR}(R_1)\neq 0$, here $R_1$ being a root of \eqref{dVeff} when $T=0$, which is the case, for instance, 
outside the compact support of a perfect-fluid object, or asymptotically in time, in a Friedmann-Robertson-Walker (background) cosmology \cite{jaimeCosmologyRevisited2012,jaimeRobustApproachGravity2011}. Then the potential has the expression 
\begin{equation}
\label{eq::JPS_Potential}
    V^{\text{eff}}(R,T) = \frac{\kappa T R}{3}-\frac{Rf(R)}{3} + \int^R f(x)dx,
\end{equation}
which can be useful to track its critical points that determine the asymptotic values for the Ricci scalar (e.g. asymptotically de Sitter or asymptotically Minkowski spacetimes, depending if the critical points correspond to 
$R> 0$ or $R=0$, respectively).

\subsection{Static and spherically symmetric (SSS) spacetimes}
\label{sec::f_R_Static_Spheerical}
Since we are interested in astrophysical situations such as the Sun and {\it nonrotating} neutron stars in $f(R)$ gravity,\footnote{The Sun and most neutron stars observed so far are rotating. However,
in this paper we do not intend to reproduce their rotation frequency but rather compare 
with objects whose rotation has a low impact on their bulk properties and where rotation is unimportant to this respect.} we consider for simplicity SSS spacetimes as a good approximation. These spacetimes can be described by the following metric,
\begin{equation}
\label{eq::f_R_Spherically_line_element}
      ds^2 =-n(r)dt^2+m(r)dr^2+  r^2 d\Omega^2,
\end{equation}
where $d\Omega^2 = d\theta^2+\sin^2\theta d\varphi^2$. From this metric and Eq. \eqref{eq::fR_formalism_3} we find a second order differential equation for the Ricci scalar
\begin{equation}
\label{eq::eq::Rpprime_1}
        R''= \frac{1}{3f_{RR}}\left[m(\kappa T+ 2f-Rf_R)-3f_{RRR}R'^2\right] + \left(\frac{m'}{2m}-\frac{n'}{2n}-\frac{2}{r}\right)R',
\end{equation}
where $'=d/dr$. From the $t-t$, $r-r$ and $\theta - \theta$  components of Eq. \eqref{eq::fR_formalism_4} and using Eq. \eqref{eq::eq::Rpprime_1} we obtain \cite{jaimeRobustApproachGravity2011},
\begin{subequations}
\label{eq::F_R_TOV_Equations}
\begin{align}
  \label{eq:mprime_1}
     m'= & \frac{m}{r(2f_{R} + rR'f_{RR})} \bigg\lbrace 2f_{R} (1-m)- 2 m r^2\kappa T^t_t  + \frac{mr^2}{3} \left(Rf_{R}+ f+2\kappa T\right), \\\notag
     & + \frac{rR'f_{RR}}{f_{R}}\left[\frac{mr^2}{3}(2Rf_{R}-f+ \kappa T) - \kappa m r^2(T^t_t+T^r_r)+2(1-m)f_{R}+ 2rR'f_{RR}\right]\bigg\rbrace, \\
      \label{eq:nprime_1_1}
     n' =& \frac{n}{r(2f_{R} + rR'f_{RR})} \left[mr^2(f-Rf_{R}+ 2\kappa T^r_r)+ 2f_{R}(m-1)- 4 rR'f_{RR}\right],\\
    \label{eq:nnprime_1_1}
      n'' =& \frac{2nm}{f_{R}}\left[\kappa T^\theta_\theta -\frac{1}{6}(Rf_{R}+ f +2\kappa T)+\frac{R'f_{RR}}{m}\right] + \frac{n}{2r}\left[2\left(\frac{m'}{m}-\frac{n'}{n}\right)+ \frac{n'r}{n}\left(\frac{m'}{m}+\frac{n'}{n}\right)\right].
\end{align}
\end{subequations}

Note that Eqs. \eqref{eq:nprime_1_1} and \eqref{eq:nnprime_1_1} are not independent. One has the freedom to choose any of the two and then use the remaining one to check the consistency of the solutions.  On the other hand, the matter variables are governed by the conservation equation $\nabla_a T^{ab}=0$, which is consistent with Eq. \eqref{eq::fR_formalism_1}, due to the diffeomorphism invariance of the theory. For a perfect fluid of 
total energy density $\rho c^2$ and pressure $p$, the EMT is $T_{ab}= (\rho c^2 + p )u_au_b + p g_{ab}$ and for the SSS scenario, the conservation equation for the EMT leads to, 
\begin{equation}
\label{eq:: TOVfR_Gravity}
    p' = - (\rho c^2 + p) \frac{n' }{2n}. 
\end{equation}
 It is important to note that $n'$ is given explicitly by the right-hand side  of Eq. \eqref{eq:nprime_1_1}, thus, Eq. \eqref{eq:: TOVfR_Gravity} is a modified version of the Tolman-Oppenheimer-Volkoff equation of hydrostatic equilibrium for $f(R)$ gravity. When $f(R)=R-2\Lambda$ one recovers from Eqs. \eqref{eq:mprime_1}--\eqref{eq:: TOVfR_Gravity} the standard equations of GR for SSS spacetimes.
 Finally, in order to close the system of equations an EOS for the fluid $p = p(\rho)$ is required. We assume two different models for the EOS: an \textit{incompressible fluid} for the Sun (cf. Sec. \ref{sec:solarsys}) where the density is taken as a step function (a non-zero constant average density inside the Sun and zero outside), and for neutron stars we take piecewise polytropic EOS and also an incompressible fluid (cf. Secs. \ref{sec:nsrhocons} and  \ref{sec:polytrop}). 

 An important quantity that can be computed from the metric is the {\it mass function} $\mu(r):$
 \begin{equation}
 \label{massfunct}
     G \mu(r)/c^2= \frac{r(m(r)-1)}{2m(r)},
 \end{equation}
 which amounts to define $m(r)= \left(1-\frac{2G \mu(r)}{c^2 r}\right)^{-1}$. The total gravitational mass of the star for asymptotically flat spacetimes corresponds to $M:=\mu(\infty)$.\footnote{When dealing with spacetimes endowed with a nonzero cosmological constant the mass $M$ is defined as 
 $M= \mu(r_*)- c^2 \Lambda r^3_*/6 G$, where $r_*$ is taken as the cosmological horizon when $\Lambda >0$ and $r_*=\infty$ when $\Lambda < 0$, such that asymptotically $m(r_*)= \left(1-\frac{2G M}{c^2 r_*}- \Lambda r_*^2/3 \right)^{-1}\sim \left(1- \Lambda r_*^2/3 \right)^{-1}\rightarrow +\infty$ (for $\Lambda >0$) and $m(r_*)\sim \left(-\Lambda r_*^2/3 \right)^{-1}\rightarrow 0$ (for $\Lambda <0$) .}

 \subsection{Regularity and asymptotic conditions}
\label{sec:rercond}

Setting  $R'=Q$, and $n'=W$, Eqs. \eqref{eq::eq::Rpprime_1} and \eqref{eq::F_R_TOV_Equations} can be written as a system of first-order ordinary differential equations
of the form $dy^i/ dr = \mathcal{F}^i(r, y^i)$ where $y^i= R,Q,W,m,n,p$. Thus we need six ``initial" conditions at the center of the star, some of which are obtained from the regularity conditions at $r=0$. This implies the following expansions near the 
center \footnote{By fixing $n(0)=1$ we find that in the asymptotic flat 
region $n(r\gg R_s)=n_\infty \neq 1$, where $R_s$ stands for the radius of the star. So at the end of the numerical integration we normalize $n(r)\rightarrow n(r)/n_\infty$ so that the {\it new} $n(r)$ is such that $n(0)=1/n_\infty$ and $n(r\gg R_s)= 1$.}
\begin{subequations}
\begin{align}
    m(r)& = 1 + m_2r^2 + \mathcal{O}(r^4),\\
    n(r)& = 1 + n_2r^2  + \mathcal{O}(r^4), \\
    R(r)& = R_0 + R_2r^2 + \mathcal{O}(r^4).
\end{align}
\end{subequations}
Hence $R'=  m'=  n'= 0$ at $r=0$. These conditions also imply, 
via Eq. \eqref{eq:: TOVfR_Gravity}, $p'(0)=0$. 
From Eqs. \eqref{eq::eq::Rpprime_1} and \eqref{eq::F_R_TOV_Equations} one finds
\begin{subequations}
    \begin{align}
    \label{Rbiprime0}
        R_2 &= \frac{\kappa T^0 + 2f^0 - R_0f_R^0}{18f_{RR}^0},\\
       n_2&=\frac{9 \kappa T^{\theta 0}_{\theta}  - 3\kappa T_{t 0}^{t} - (R_0 f_R^0 + f^0 + 2\kappa T^0)}{18\,f_{R}^0},\\
       m_2& = \frac{R_{0}f_{R}^0+f^{0}+2\kappa T^0-6\kappa T^{t0}_{t}}
           {18\,f_{R}^0},
    \end{align}
\end{subequations}
where the {knotted} quantities indicate evaluation at $r=0$. 
For the cosmologically motivated $f(R)$ models (see Sec. \ref{Sec::Viablef_Rmodels} below), it is expected that the spacetime outside the star is asymptotically Schwarzschild-de Sitter. This asymptotic behavior implies that $R\rightarrow R_{1}$ as $r\rightarrow \infty$ such that $R_1$ coincides with the critical point where $dV^{\text{eff}}(R,0)/dR= (2f- f_RR)/3$ vanishes. Nevertheless, for other $f(R)$ models (see Sec. \ref{sec:starsR2}), $R_1$ could be zero, in which case the spacetime becomes asymptotically flat. In order to obtain the correct asymptotic value, one needs to choose the adequate value for the Ricci scalar at the center of the star $R_0$. 
This boundary value problem (BVP) can be treated as an initial value problem (IVP) by imposing the regularity conditions at the center of the star ($r=0$) together with the values of the matter 
sources there (like the density and pressure)  with the aid of a shooting method in such a way that an adequate value for $R_0$ leads to the correct asymptotic value $R_1$. For DE models one usually demands $R\rightarrow R_{1}\neq 0$ asymptotically so that $\Lambda_{eff}= R_1/4$. This effective cosmological constant is the one that emerges naturally in a cosmological setting when one evolves the system in cosmic time from the past (e.g. from the radiation-dominated epoch) to the far future \cite{jaimeCosmologyRevisited2012}. Now, when solving the problem numerically for the Sun, the scales where the asymptotic value $R_1$ is reached are cosmological distances, which are far beyond the solar system, and therefore are almost impossible to reach in actual computations. So for all practical purposes, the spacetime around the Sun should look like a perturbation of Minkowski spacetime with $n(r)\times m(r)\sim 1$, in order to pass the usual tests. But $f(R)$ models that fail to develop the chameleon or screening mechanism usually exhibit a quotient $(m(r)-1)/(1-n(r))\sim 1/2$ outside the star, even if $n(r)$ and $m(r)$ approach $1$ when $R_\odot \ll r$. The viable $f(R)$ models in the solar system are those for which $(m(r)-1)/(1-n(r))\sim 1$ when $R_\odot \ll r$. This implies that the post-Newtonian parameter $\gamma$ takes 
the value $\gamma \sim 1/2$ for non-viable models and $\gamma\sim 1$ for the viable ones. 
The inner domain is defined as $r\in [0, R_s]$, where $R_s$ is the radius of the star fixed at the place where the pressure vanishes. We emphasize that outside this inner domain the solution is (in general) not the Schwarzschild solution since the 
Ricci scalar contributes as a kind of {\it geometric matter} due to the non-vanishing 
scalar-degree of freedom outside the star. However, if the screening mechanism ensues, this contribution should be small enough in order for the exterior solution for $R(r)$ to be suppressed in the neighborhood of the solar system, and then this solution should be very close to the Schwarzschild solution.
We use a Runge-Kutta algorithm of {\it fifth} order to solve the set of 
differential equations together  with an arbitrary-precision arithmetic in a code written in \texttt{JULIA}. Additional details about the dimensionless form of the equations and the numerical implementation can be found in Appendix \ref{secc::Numerical_Strategy}.
\subsection{Viable (cosmological) $f(R)$ models}
\label{Sec::Viablef_Rmodels}
One of the motivations to study  $f(R)$ gravity is that it can account for the late accelerated cosmic expansion and at the same time be distinguished from the $\Lambda {\rm CDM}$ model by predicting an EOS that evolves in cosmic time.  However, certain criteria on the mathematical expression for $f(R)$ must be imposed in order to obtain a consistent cosmological evolution. In this section we introduce some of the $f(R)$ models, that together with the matter sector, mimic  the dark matter and dark energy epochs correctly (that follows the radiation dominated era at early times). 
All these models have the feature that they do not include an explicit cosmological constant, in the sense that $f(0) = 0$, in contrast to what happens in GR ($f_{GR}(R) = R - 2\Lambda$) where $f_{GR}(0)= -2 \Lambda$. In particular, for an $f(R)$ model that is able to produce dynamically an effective cosmological constant $\Lambda_{eff}$, the Ricci scalar must approach asymptotically a constant value $R_1=4 \Lambda_{eff}$ which coincides precisely with a minimum of a ``potential'' $V^{\text{eff}}(R,0)$, defined earlier via Eq. \eqref{dVeff}. 
This quantity appears precisely in the r.h.s of Eq. \eqref{eq::fR_formalism_3}. Regarding stability, it is common to impose the conditions $f_{RR}>0$ and $f_{R}>0$ (see \cite{jaimeCosmologyRevisited2012} for a thorough discussion). 

On the other hand, $f(R)$ gravity must also satisfy local gravity constraints such as the solar-system tests. In particular, a careful analysis of $f(R)$ models in that context 
 is found in several studies \cite{chibaGravityScalarTensorGravity2003,chibaSolarSystemConstraints2007,negrelliSolarSystemTests2020,guoSOLARSYSTEMTESTS2014,capozzielloSolarSystemEquivalence2008, liuConstrainingGravitySolar2018,faulknerConstrainingGravityScalartensor2007,delacruz-dombrizCommentViableSingularityFree2009, zhangBehaviorGravitySolar2007,negrelliSolarSystemTests2020}, which, in contrast to the current work, use approximations in the equations. Here we focus on models which, in general, can be easily seen as the usual Ricci scalar $R$ term of GR  plus some function of $R$ that has a built-in scale of the order of $H_0^2$ (where $H_0$ is the Hubble constant) in order to produce the correct cosmological history. This means that at high curvatures, where $\vert R\vert \gg  H_0^2/c^2$, the model acquires the general form $f(R)\approx R- \Lambda_{\text{eff}}^\infty$ where $\Lambda_{\text{eff}}^\infty$ is another effective cosmological constant also of the order of $H_0^2/c^2$ (not to be confused with the low curvature regime ($R= R_1\sim H_0^2/c^2$) where $\Lambda_{eff}=R_1/4$). Thus, some of these models are carefully built to follow a GR type curvature when the matter content has large densities (compared with the cosmological ones) such as in stellar configurations. A class of models that meet these requirements were proposed by Hu-Sawicki \cite{huModelsCosmicAcceleration2007} and Starobinsky \cite{starobinskyDisappearingCosmologicalConstant2007}. We also include the logarithmic $f(R)$ model proposed by \citet{mirandaViableSingularityFreeGravity2009} to test its viability with respect to stellar configurations.
\subsubsection{Starobinsky $f(R)$ model}
\label{sec::Starobinskyf(R)model}
This model is defined by \cite{starobinskyDisappearingCosmologicalConstant2007}:
\begin{equation}
\label{eq::Starobinskymodel}
    f(R)_S= R + \lambda R_S\left[\left(1+\frac{R^2}{R_S^2}\right)^{-q}-1\right],
\end{equation}
with $q$ and $\lambda$ positive dimensionless parameters and $R_S$ a cosmological scale given by $R_S = \sigma_S H_0^2c^{-2}$, where $\sigma_S$ is an order-one dimensionless parameter. Some choices have been considered in recent works \cite{devDelicateGravityModels2008}. Here we use $\lambda=1$, $\sigma_s=2$, and $q=4.17$ as in Refs. \cite{motohashiGravityItsCosmologicals2011, jaimeCosmologyRevisited2012} where it 
is demonstrated that the Starobinsky model is able to produce an adequate matter epoch prior to a successful accelerated era. 
Static and spherically symmetric configurations under the Starobinsky model \eqref{eq::Starobinskymodel} have been studied numerically and non-perturbatively in Refs. \cite{upadhyeExistenceRelativisticStars2009, babichevRelativisticStarsGravity2009, babichevRelativisticStarsScalartensor2010a} using the STT approach, and a semi-linear analysis to treat the solar system tests is presented in \cite{negrelliSolarSystemTests2020}.  

\subsubsection{Hu-Sawicki $f(R)$ model}
\label{eq::Hu_Sawcki_model}
This model takes the following form \cite{huModelsCosmicAcceleration2007}:
\begin{equation}
\label{eq::Hu_Sa_Model}
f(R)_{\text{HS}} =     R - \frac{c_{1} m^{2} \left(\frac{R}{m^{2}}\right)^{n}}{c_{2} \left(\frac{R}{m^{2}}\right)^{n} + 1},
\end{equation}
where $c_1$, $c_2$ and $n>0$ are free dimensionless parameters. The constant $m^2$ is defined as $m^2 := k^2\bar{\rho}_0/3$, with $\bar{\rho}_0$ being the average matter density of the universe today. We take the numerical value $m^2 = 0.24H_0^2c^{-2}$ \cite{jaimeCosmologyRevisited2012,huModelsCosmicAcceleration2007}. 
A suitable choice of the free parameters is given in Refs. \cite{jaimeRobustApproachGravity2011,negrelliSolarSystemTests2020}, $c_1 = 1.25\times10^{-3}$, $c_2 = 6.56 \times 10 ^{-5}$ and $n=4$, in order to achieve a late-time cosmological acceleration compatible with observations. Notice that this $f(R)$ model can be written as a sum of three terms: 
\begin{equation}
\label{eq::Hu_Sa_Model_1}
f(R)_{\text{HS}}= R - \frac{c_1 m^2 }{c_2} + \frac{c_1 m^2}{c_2\left[1 +  c_2 \left(\frac{R}{m^2}\right)^n\right]}.
\end{equation}
At high curvatures where $R\gg m^2$, the third term of \eqref{eq::Hu_Sa_Model_1} vanishes and it reduces to $f(R)_{\text{HS}}\approx R  - c_1 m^2/c_2$. On the other hand, at low curvatures, all three terms of \eqref{eq::Hu_Sa_Model_1}  are of the same order, and the model separates itself from GR. 
\subsubsection{MJWQ $f(R)$ model}
\label{secc_MJW_MODEL}
 The following logarithmic model was proposed by \citet{mirandaViableSingularityFreeGravity2009}: 
\begin{equation}
    \label{eq::MirandaModel}
    f(R)_{\text{MJWQ}} = R -\alpha_{M} R_*\ln{\left(1+ \frac{R}{R_*}\right)}, 
\end{equation}
where $R_*$ and $\alpha_{M}$ are free positive parameters. While this model can correctly reproduce the cosmological history at the background level \cite{jaimeCosmologyRevisited2012}, 
such a model seems to be inconsistent with the power matter spectrum and with the solar system tests \cite{bisabrSolarSystemConstraints2010,delacruz-dombrizCommentViableSingularityFree2009, mirandaMirandaReply2009,negrelliSolarSystemTests2020}. Here we use $R_*=H_0^2/c^2$,
and $\alpha_M=2$ as in Refs. \cite{mirandaViableSingularityFreeGravity2009,jaimeRobustApproachGravity2011} and try to reanalyze the solar system tests under the non-perturbative set of equations discussed in Secs. \ref{sec::f_R_Static_Spheerical} and \ref{sec:rercond}.


\section{Solar system tests}
\label{sec:solarsys}

Although, the Schwarzschild–de Sitter (SdS) metric is a vacuum solution for the $f(R)$ cosmological models considered in this work and, as mentioned above, we seek an exterior solution outside the Sun to approach the asymptotic de Sitter behavior, we will see that there is no guarantee that the whole solution (the interior and the exterior ones) is physically viable, especially the exterior solution for the metric. For example, the exterior solution can be asymptotically dS, but due to the contribution of $R(r)$, the falloff rate of the metric could be slower than in GR outside the Sun. In other words, in GR 
(without $\Lambda$) $R=-\kappa T$, so in the case of a perfect fluid, the Ricci scalar automatically vanishes outside the compact support of the star (e.g. the Sun) and does not contribute to the 
solution for the metric.\footnote{Even if we include the actual value for $\Lambda$ in 
the analysis of the Sun within GR, we have that outside the Sun $R/(\kappa \rho_\odot c^2)= 4\Lambda/(\kappa \rho_\odot c^2) \sim 10^{-29}$ (cf. Table \ref{table:2}). In a similar way, 
the contribution $\Lambda r^2\sim 10^{-26}$ to the metric is very small within the solar system, and the tests on the PPN $\gamma$ parameter are not sensitive to this contribution. The contribution of $\Lambda$ within the solar system is thus negligible in this respect.} In $f(R)$ gravity the situation is different, the Ricci scalar satisfies a second order differential equation, and its value 
in the neighborhood of the star surface does not necessarily vanish 
(or becomes small) like in the GR scenario.
So we emphasize again that in order to pass the solar system tests, the exterior solution for $R(r)$ must be suppressed sufficiently by the screening mechanism inherent to the specific $f(R)$ model itself. As we remarked before, some studies of solar system tests under rather different approaches and approximations
in $f(R)$ gravity are found in the following references \cite{chibaGravityScalarTensorGravity2003,chibaSolarSystemConstraints2007,negrelliSolarSystemTests2020,guoSOLARSYSTEMTESTS2014,capozzielloSolarSystemEquivalence2008, liuConstrainingGravitySolar2018,faulknerConstrainingGravityScalartensor2007,delacruz-dombrizCommentViableSingularityFree2009, zhangBehaviorGravitySolar2007}.

As discussed in Section \ref{Sec::Viablef_Rmodels}, each $f(R)$ model is characterized by an inherent curvature scale of the order of $R_*:= H_0^2/c^2$, which can be used to define an associated mass density.
\begin{equation}
\label{eq::rho_ast}
    \rho_*:= \frac{c^4}{8\pi G}\frac{R_*}{c^2} \approx 3.008\times 10^{-27} \text{kgm}^{-3},
\end{equation}
which is of the order of the average cosmological density. On the other hand, if we take the average density of the Sun as $\rho_\odot = 1408 $kg/m$^3$, or a typical neutron star density $\rho_{\text{NS}}\backsimeq 10^{17} \text{kg m}^{-3}$, the density contrast between 
$\rho_*$ \eqref{eq::rho_ast} and these two densities is colossal. This stark contrast is shown in Table \ref{table:2}, where the ratio between the solar density $\rho_\odot$ and the characteristic cosmological density $\rho_*$ is on the order of $\rho_\odot/\rho_* \sim10^{29}$, and the disparity becomes even larger when considering neutron stars ($\rho_\text{NS}/\rho_* \sim 10^{44}$). This huge ratio appears naturally in the field equations, as seen from their dimensionless form \eqref{eq::f_R_TOV_ADIM_EQUATIONS} and the dimensionless parameters \eqref{eq::Numerical_coeficients} and has been a technical issue that one faces when integrating stellar objects embedded in a realistic de Sitter background \cite{kobayashiRelativisticStarsGravity2008,frolovSingularityProblemModels2008,upadhyeExistenceRelativisticStars2009}. 
\begin{table}[h!]
\centering
\begin{tabular}{||c c c c||} 
 \hline
 & $\rho$[ kg/m$^3$] & $\rho/\rho_\odot$ &  $\rho/\rho_*$ \\ [0.5ex] 
 \hline\hline
  $\rho_\odot$ & 1408 & 1 &  $4.6808\times 10^{29}$\\
 $\rho_*$ & $3.008 \times 10^{-27}$& $2.1363\times 10^{-30}$& 1 \\
  $\rho_\text{NS}$ & $10^{17}$& $\sim  10^{14}$& $\sim 10^{44}$\\ [1ex] 
 \hline
\end{tabular}
\caption{Fraction between the characteristic densities of the Sun and a neutron star and the cosmological density.}
\label{table:2}
\end{table}

For the purpose of the solar system tests and in order to avoid unnecessary complications associated with the EOS for the Sun, we assume a constant-density star taking $\rho=\rho_0=\rho_\odot$ as the average density of the Sun inside the object and zero outside. That is, we take a step function for Sun's density, where the radius is defined at the place where the pressure vanishes. For this simplified matter model we calibrated the central pressure starting with GR so that the final Sun's configuration results in a mass and radius close to its actual values (see Appendix \ref{app:sunGR}). Nevertheless, for $f(R)$ theories, due to the high density contrast alluded to above, we are compelled to start with densities and central pressures for the object which are much lower than the actual ones for the Sun. Then we implement a gradual ``density ramping" strategy in our numerical integration such that $\rho_0$ approaches $\rho_\odot$ gradually.  We begin with a density close to the characteristic cosmological density, and gradually increase it and try to find solutions that represent a configuration of the Sun similar to the GR solutions 
(Appendix \ref{app:sunGR}). More specifically, for the inner density we start with $ \rho_0 = \xi \rho_*$ taking $\xi=1$ and increase the value for $\xi$  until we reach the average density of the Sun. This implies an increase in $\rho_0$
about 29 orders of magnitude in relation to $\rho_*$ (cf. Table \ref{table:2}).
However, as we will illustrate below, this strategy fails even before we reach the density $\rho_\odot$. We also fixed the central pressure in terms of $\rho_0$. 
Using this methodology, we numerically solve the system of equations \eqref{eq::eq::Rpprime_1}--\eqref{eq:: TOVfR_Gravity} for each value of $\xi$ in order to monitor the numerical solutions. Note that for each $\xi$ we have to find
the adequate shooting value $R_0$ so that $R\rightarrow R_1$ asymptotically, where $R_1$ is the cosmological value determined by the specific model $f(R)$. In this way, we do not assume {\it a priori} that the object is close to the Newtonian weak field, but solve the complete system of ODEs without any approximation. We have tested this methodology for the GR scenario (without cosmological constant) and confirm that the results are basically the same (up to numerical errors associated with the Runge-Kutta discretization) to the analytic GR solution for an {\it incompressible} fluid for the Sun and to the corresponding Newtonian expectations even if the full nonlinear equations are solved numerically (see Appendix \ref{app:sunGR}). In other words, 
the system of equations automatically recognizes when the object is not relativistic and the resulting solution becomes close to a perturbed Minkowski spacetime even if the full general relativistic equations are solved numerically. The purpose of proceeding this way is because the so called {\it screening, thin-shell} or {\it chameleon} effect is supposed to be a high nonlinear mechanism that suppresses the scalar degree of freedom outside the central object (the Sun in this case), and thus, our goal is to avoid an {\it a priori} linearization in the equations that might be interpreted as an {\it ad hoc} implementation in order to precisely obtain what one is trying to find. 


\subsection{Constant-density stars in MJWQ $f(R)$ gravity}
We begin our analysis by constructing a collection of star configurations for the MJWQ $f(R)$ model \eqref{eq::MirandaModel} taking $\alpha_M=2$.  For this model, the Ricci scalar $R$ will approach the de Sitter minimum $R_1/R_*\approx 6.1461833318$ asymptotically, which corresponds to the algebraic root of the effective potential \eqref{dVeff} in the absence of matter. Our primary interest lies in solutions where the Ricci scalar $R$ follows closely the stellar density profile, that is, solutions where $R$ does not deviate too much from its GR value given by $R=-\kappa T=-\kappa(3p-\rho c^2)$ inside the star. Therefore, $R$ is expected to interpolate smoothly between the high-density regime ($R\gg R_*$) inside the star, and a low density regime ($R\sim R_*$) outside the star where by construction $\rho=0$. We explore the solutions by varying the central pressure, $p_0$, which parametrizes the star configurations. For each $p_0$ one is required to find the central value of the Ricci scalar, $R_0$, by a shooting method, so that asymptotically $R\rightarrow R_1$ which corresponds to the de Sitter value  for this model, as we stressed before.

\begin{figure}
\centering
\includegraphics[width=\linewidth]{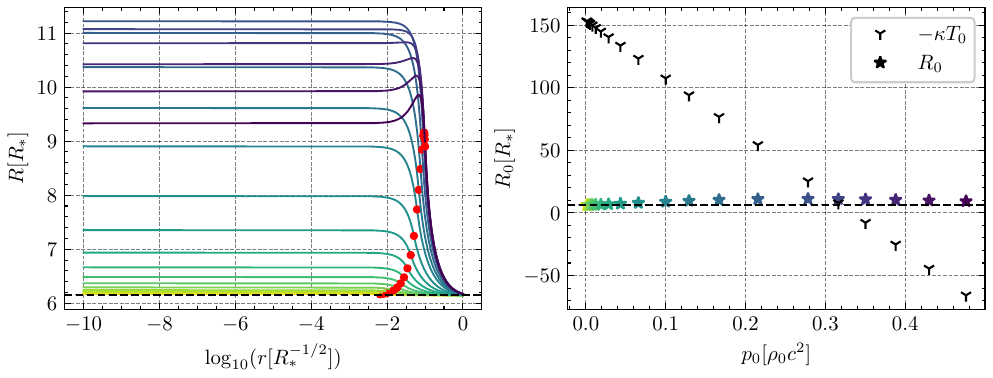}
\caption{\textit{Left panel:} Radial profiles for the Ricci scalar for different values of the central pressure $p_0$. The red dots indicate the location of the star's radius where the pressure vanishes. 
Colors from green to violet indicate increasing central pressures, from $p_0=10^{-3}\rho_0 c^2$ to $p_0=0.5\rho_0 c^2$.
\textit{Right panel:} Ricci scalar at 
$r=0$ (associated with the solutions depicted in the left panel) as a function of $p_0$. 
The plots belong to the MJWQ model with $\alpha_M=2$ and fixed density $\rho_0=25R_1/(\kappa c^2)$. In these units, the asymptotic de Sitter value is $R_1/R_*\approx 6.1461833318$ (horizontal black dashed lines)
where $R_*:= H_0^2/c^2$.}
\label{FIG::MJWQ_ALPHA_1_2_multiple_presusre}
\end{figure}

Left panel of Fig. \ref{FIG::MJWQ_ALPHA_1_2_multiple_presusre} depicts different solutions for the Ricci scalar $R$ for a low-density star with $\rho_0=50 R_1c^2/(16\pi G)$. This density is still several orders of magnitude below the realistic value for the Sun's density.  As the central pressure $p_0$ increases  (ranging from $p_0=10^{-3}\rho_0 c^2$ to $p_0=0.5 \rho_0 c^2$ and colors from green to violet), 
the Ricci scalar $R(r)$ shows a characteristic pattern where instead of 
decreasing monotonically from $R_0$ to the asymptotic value $R_1$, it increases and after reaching a maximum 
value inside the star, it decreases monotonically to $R_1$ (cf. the plots in violet). 
The Ricci scalar $R_0$ also increases with $p_0$, reaches a maximum value, and then decreases as one can observe more clearly from the right panel of Fig. \ref{FIG::MJWQ_ALPHA_1_2_multiple_presusre}. For comparison, the corresponding GR value $R_0=-\kappa T_0$ for each configuration is also displayed (right panel) in conjunction with the values $R_0$ for the MJWQ model, both as a function of the central pressure $p_0$. For low central pressures (greener colors) $R_0$ deviates the most from its GR counterpart, while for higher central pressures $R_0$ approaches the latter but then deviates again. 

\begin{figure}
\centering
\includegraphics[width=\linewidth]{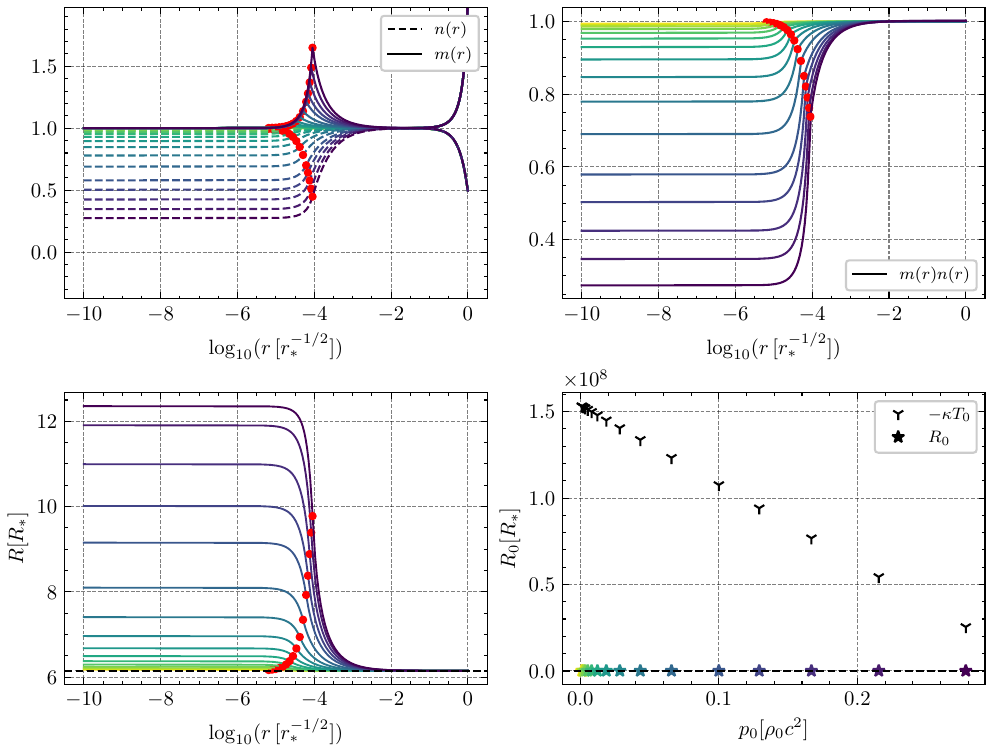}
\caption{Solutions for the MJWQ $f(R)$ model. \textit{Top left panel:} Metric potentials $n(r)$ (dotted lines) and $m(r)$ (solid lines) for a star with  $\rho_0= 10^7 R_1/(2\kappa c^2)$  and central pressures from $p_0=10^{-3}\rho_0 c^2$ to $p_0=0.3\rho_0 c^2$ (green to violet colors). The red dots indicate the location of the star's surface where the pressure vanishes. The metric functions show the de Sitter behavior asymptotically $n\sim 1 - \Lambda_{eff} r^3/3$ and $m\sim n^{-1}$. \textit{Top right panel:} Product of the metric potentials $m(r)\times n(r)$ for the same configurations. Contrary to GR, the product deviates from one outside the star.  \textit{Bottom left panel}: Radial profiles for the Ricci scalar. \textit{Bottom right panel}: Ricci scalar at the center of the star (associated with the solutions at
the bottom left panel) as function of the central pressure (color stars).
The corresponding GR values $-\kappa T_0= -\kappa (3p_0-\rho_0 c^2) $ (\texttt{Y}) are shown for reference.}
\label{FIG::potentialsMJW_Multiple_Pressure_for_rho_5.00000e+07}
\end{figure}

For this model we could explore higher densities and Fig. \ref{FIG::potentialsMJW_Multiple_Pressure_for_rho_5.00000e+07} shows 
solutions for a star with density $\rho_0=10^7R_1c^2/(2\kappa c^2)$ and central pressures ranging from $p_0=10^{-3}\rho_0 c^2$ to $p_0=0.3 \rho_0 c^2$. Our results are compatible with those found in 
Refs. \cite{kobayashiCanHigherCurvature2009, jaimeRobustApproachGravity2011} where 
similar configurations were analyzed. Nevertheless, the solutions are not only quite 
different from the GR expectations but they become increasingly difficult to find as the density 
becomes larger. So, we could not reach the solar density to check if the MJWQ model 
is or is not able to satisfy the PPN value $\gamma\sim 1$ outside the Sun.

\subsection{Constant-density stars in Starobinsky $f(R)$ gravity}

We now turn our attention to the Starobinsky $f(R)$ model and conduct an analysis similar to the 
MJWQ model. We use the parameters described in \ref{sec::Starobinskyf(R)model} which implies a de Sitter point at 
$R_1/R_*\approx 6.82795$.

\begin{figure}
\centering
\includegraphics[width=\linewidth]{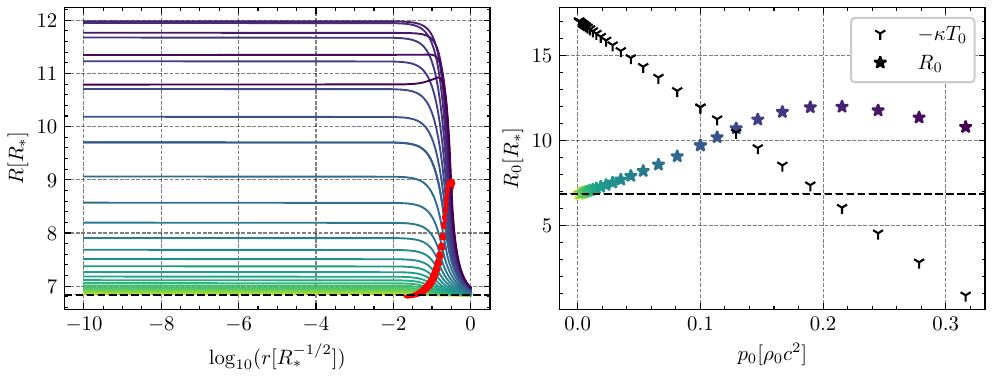}
\caption{{\it Left panel:} Ricci scalar for the Starobinsky model with central pressures 
from $p_0=10^{-3}\rho_0 c^2$ to $p_0=0.3\rho_0 c^2$ (colors from green to violet) 
and fixed density $\rho_0= R_1/(4\kappa c^2)$. The asymptotic de Sitter value is $R_1/R_*\approx 6.82795191$, where $R_*:= H_0^2/c^2$. 
{\it Right panel:} the central values $R_0$ associated with the solutions of the left panel (color stars)
plotted with respect to the central pressures $p_0$. For reference the corresponding GR values 
$-\kappa T_0= -\kappa (3p_0-\rho_0 c^2) $ are shown (\texttt{Y}).} 
\label{FIG::STBS_JAIME_MultiplePressure_0.5}
\end{figure}

\begin{figure}
\centering
\includegraphics[width=\linewidth]{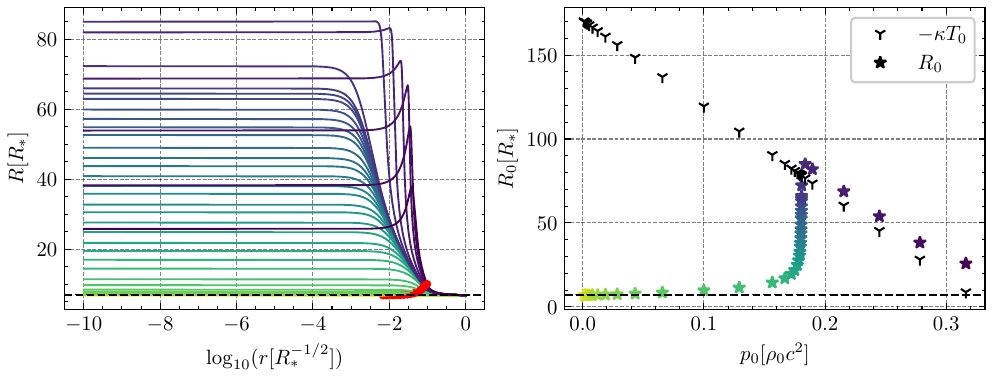}
\caption{Similar to Fig.\ref{FIG::STBS_JAIME_MultiplePressure_0.5} but taking $\rho_0= 25R_1/(\kappa c^2)$.} 
\label{FIG::STBS_JAIME_MultiplePressure_50}
\end{figure}

Left panel of Fig.  \ref{FIG::STBS_JAIME_MultiplePressure_0.5} shows the solutions for $R$ 
for different $p_0$, from $10^{-3}\rho_0 c^2$ to $0.3\rho_0 c^2$ with a fixed density $\rho_0=R_1/(4\kappa c^2)$. 
The results are similar to the MJWQ model, however, in the Starobinsky model the dependence of these solutions
with respect to $p_0$ is more pronounced, exhibiting a sharper transition from the star's interior to the exterior. Figure \ref{FIG::STBS_JAIME_MultiplePressure_50} illustrates this behavior more clearly for a higher density star with $\rho_0= 25R_1/(\kappa c^2)$. In this case a narrow region emerges (right panel) where small variations in $p_0$ result in substantial changes in $R_0$. These are precisely the kind of configurations that we would expect to match the GR solutions: $R$ interpolates between two distinct curvature regimes like a step function. Notwithstanding, in this scenario the solutions are associated with higher central pressures, whereas a Sun configuration 
in GR is associated with $p_0\ll \rho_0 c^2$.

We explored higher densities, but like in the MJWQ model, a significant increase in the numerical precision was needed 
to find a solution with the right value $R_0$ giving the correct behavior $R(r)\rightarrow R_1$ asymptotically. 
Moreover, such solutions required also high central pressures. For low pressures, essential for an adequate 
model for the Sun, the solutions $R(r)$ that we explored were systematically close to the de Sitter value $R_1$ even 
at the interior of the star, instead of being correlated with the density. The right panel of Fig.
\ref{FIG::STBS_JAIME_MultiplePressure_50} shows that $R_0\ll R_0^{GR}=-kT_0$ when $p_0\ll \rho c^2$, and therefore, the central values of $R_0$ in the Starobinsky model are not close to the corresponding values in GR. 

In summary, for the Starobinsky model we were also unable to check if the solar system tests are or are not 
verified as we could never get close to a star with $\rho\sim \rho_\odot$ and $p_0\ll \rho c^2$ since the numerical 
approach failed long before reaching that density and low pressures.

\subsection{Constant-density stars in Hu-Sawicki $f(R)$ gravity}
\begin{figure}
\centering
\includegraphics[width=\linewidth]{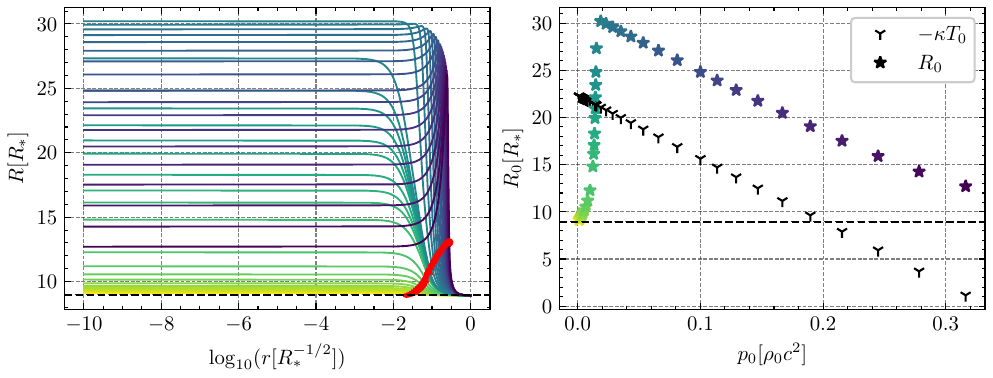}
\caption{{\it Left panel:} Ricci scalar associated with
solutions for the Hu-Sawicki $f(R)$ model at a fixed density  $\rho_0= R_1/(4\kappa c^2)$
and central pressures ranging from $p_0=10^{-3}\rho_0 c^2$ to $p_0=0.3\rho_0 c^2$ (colors from green to violet indicate increasing central pressures). The de Sitter minimum is at $R_1/R_*\approx 8.931080$ (dashed line). The red dots indicate the location of the star's surface. {\it Right panel:}. Ricci scalar at the center of the star (corresponding to solutions of left panel) as a function of the central pressure (color stars). Like in previous figures, the GR value 
is plotted for reference (\texttt{Y}).}
\label{fig::HUSAW_Multiple_Pressure_for_rho_5.00000e+00}
\end{figure}

\begin{figure}
\centering
\includegraphics[width=\linewidth]{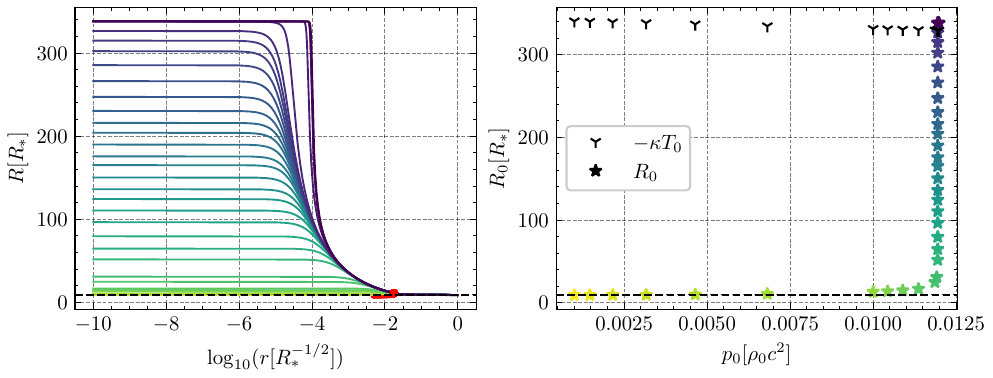}
\caption{Similar to Fig.\ref{fig::HUSAW_Multiple_Pressure_for_rho_5.00000e+00} but taking 
 $\rho_0= 50 R_1/(\kappa c^2)$ and central pressures ranging from $p_0=10^{-3}\rho_0 c^2$ to $p_0=0.0125\rho_0 c^2$.  }
\label{fig::2Alt_HUSAW_Multiple_Pressure_for_rho_1.00000e+02}
\end{figure}

For this $f(R)$ model, which has a structure similar to the Starobinsky model, the global minimum of its effective potential is at $R_1/R_*\approx 8.93108$. 
Left panel of Fig. \ref{fig::HUSAW_Multiple_Pressure_for_rho_5.00000e+00} depicts the profiles of Ricci scalar for a star of density $\rho_0= R_1/(4\kappa c^2)$ for increasing central pressures ranging from $10^{-3}\rho_0 c^2$ to $0.3\rho_0 c^2$. As in the Starobinsky model, we observe the emergence of a characteristic screening: the higher the central pressure, the steeper the gradient of the solution $R(r)$. This behavior indicates that the high curvature values are 
confined mostly within the star and the gradients in a narrow region near the surface of the star. We considered also a higher density star with $\rho_0=50 R_1/(\kappa c^2)$ (see Fig. \ref{fig::2Alt_HUSAW_Multiple_Pressure_for_rho_1.00000e+02}). For this higher density and within a narrow range of  central pressures between $0.01\rho_0 c^2$ and $0.0125\rho_0 c^2$, a sudden increase of $R_0$ happens, reaching values close to the GR value $\kappa (\rho_0 c^2-3p_0)$. However, starting at this density 
and for higher central pressures, it proves to be particularly challenging 
to obtain solutions since a higher 
numerical precision for $R_0$ is needed to achieve a successful {\it shooting} to the desired asymptotic value $R_1$. 
Due to this difficulty, we could only find solution for low pressures, but the solutions for $R(r)$ remain close to $R_1$ 
instead of tracking the values of the density inside the star. 

\section{Neutron stars in R-squared gravity (non cosmological $f(R)$ models)}
\label{sec:starsR2}
Unlike the $f(R)$ models presented in Section \ref{Sec::Viablef_Rmodels} to explain a 
{\it late} cosmic accelerated expansion, the quadratic model was first proposed by 
Starobinsky as a model of inflation for the early universe \cite{starobinskyNewTypeIsotropic1980}. It is given by the following function
\begin{equation}
\label{eq::f_R_RSquered_model}
    f(R) = R + a R^2,
\end{equation}
where $ a >0$ is a parameter with dimensions of {\it length squared}. This model is part of a wider class of $f(R)$ models where $f(R)= R+ \alpha R^n$ ($\alpha> 0$ and $n>0$) \cite{defeliceFRTheories2010}. Additionally, there are other similar higher-order models such as $f(R)= R + \alpha R^2 + \beta R^3$ or $f(R)=R + \alpha R^2(1 + \gamma \ln (R/R_0))$ \cite{capozzielloMassRadiusRelationNeutron2016}.  Although some of these models emerged to achieve inflation, many authors have implemented them to the presence of strong gravitational fields. 
In particular, the model \eqref{eq::f_R_RSquered_model} (hereafter referred to as $R-squared$ gravity) 
was adapted to be tested in NS by changing the value of $a$ (from the value used to describe the inflationary epoch) and adjust it to have length-scales of the order of the scales found in those stars, i.e., $a$ ranging from $\sim {\rm km}^2$ to $\sim 10^3 {\rm km}^2$. The idea is then to 
analyze the deviations from GR in this scenario \cite{cooneyNeutronStarsGravity2010, capozzielloMassRadiusRelationNeutron2016, arapogluConstraintsPerturbativeGravity2011,  astashenokRealisticModelsRelativistic2017}.
 In this way, the parameter $a$ for the $R-squared$ model Eq. \eqref{eq::f_R_RSquered_model} determines not a cosmological scale 
 but a typical scale for a compact stellar object. 

This scale already reveals a stark difference with respect to the characteristic curvature scale of the order of $R_*= H_0^2/c^2$ of viable dark energy $f(R)$ models described in the previous section, given that $ a \ll 1/R_*$. Due to this feature, it has been argued that this class of $f(R)$ models (with much shorter length scales) can pass the
solar system tests comfortably, since the relevant term at these scales is the linear one in
$R$ \cite{capozzielloMassRadiusRelationNeutron2016}.

The argument is simply as follows: $f(R)= R + a R^2= R(1+ aR)$. The
Ricci scalar inside the Sun according to $GR$ is $R\sim \kappa \rho_\odot c^2 \sim 10^{-3}/r_\odot^2$. Considering that in this $f(R)$ model the deviations with respect to GR within the solar system are small, then within the Sun $f(R)\sim R [ 1 + a/(10^3 r_\odot^2)]$. Taking $a\sim {\rm km}^2$ and $r_\odot \sim 0.7 \times 10^{6} {\rm km}$, then $a/(10^3 r_\odot^2)\sim 10^{-15}\ll 1 $. Even if one takes
$a\sim 10^3 {\rm km}^2$ the non-linear contribution remains very small. Since outside the Sun one assumes that the Ricci scalar is highly suppressed, the non-linear contribution becomes even smaller in the solar system. Although these
arguments seem reasonable, if one puts forward this model as a {\it universal} gravity theory, such estimates should be checked carefully within the solar system without {\it a priori} 
assuming their validity, but as emerging naturally from the numerical integration of the differential equations of the model itself.
We now focus on the analysis of static and spherically symmetric NS in $R-squared$ gravity. In recent years, this kind
of study has been approached in the literature in two main ways: the first implements a perturbative method around GR, and the second computes NS configurations from the solution of the complete set of differential equations of $f(R)$ gravity. For instance, in Refs. \cite{cooneyNeutronStarsGravity2010,arapogluConstraintsPerturbativeGravity2011,orellanaStructureNeutronStars2013} 
the authors follow a perturbative approach, and in particular, 
in Refs. \cite{arapogluConstraintsPerturbativeGravity2011,orellanaStructureNeutronStars2013}, mass-radius curves are reported in order to put bounds on the value of $a$ for a sample of EOS. Nevertheless, in those perturbative approaches, the authors do not establish the range of validity of their approximations relative to the
full non-perturbative analysis where one does not assume $aR\ll 1$, and where the full theory and the full differential set of equations are taken into account.

On the other hand, in Ref. \cite{yazadjievNonperturbativeSelfconsistentModels2014} 
a non-perturbative method is developed, but it resorts to the STT transformation of $f(R)$ gravity in the Einstein frame and the authors solve numerically the set of differential equations in that frame for four realistic EOS. They argue precisely about the {\it inconsistency and lack of reliability} of the perturbative approach. Similarly, \citet{kaseNeutronStarsGravity2019} presented numerical solutions using the STT, both in the Jordan and Einstein frames for different class of theories. 
In the case of {\it R-squared} model, their method did not allow them to find NS solutions with the 
appropriate asymptotic Minkowski behavior, in contrast with our results and those of Ref.\cite{yazadjievNonperturbativeSelfconsistentModels2014}\footnote{\citet{kaseNeutronStarsGravity2019} stress the following: {\it 
We performed numerical simulations in the
Starobinsky model by varying $\phi_0$ at r = 0 and did not ﬁnd consistent solutions satisfying all the boundary conditions
discussed above for both SLy and FPS EOS. The analysis of Ref. [28] based on the polytropic EOS...also
reached the same conclusion...we showed that BD theories with a 
positive constant mass squared $m^2$ {\rm ($m^2=1/6a$)} generally face the problem of realizing stable NS field configurations consistent with all the boundary conditions. Apart from ref.[28], this fact was overlooked in most of the past works about NS solutions in $f(R)$  gravity with the positive constant $m^2$ outside the star.}  This comment is in conflict with the findings presented in Ref. \cite{yazadjievNonperturbativeSelfconsistentModels2014}, where the authors do not find such {\it unstable} field 
configurations, despite using a similar STT approach in the Einstein frame. Kase \& Tsujikawa, who cites \citet{yazadjievNonperturbativeSelfconsistentModels2014} themselves, do not 
try to argue or clarify the origin of this discrepancy and conflict. As we will comment later in 
Sec. \ref{sec:polytrop} we do not find those {\it instabilities} either and our results are consistent with 
Ref. \cite{yazadjievNonperturbativeSelfconsistentModels2014}.}.
Other studies of NS in $R-squared$ gravity using the STT approach are found in 
Refs.~\cite{staykovSlowlyRotatingNeutron2014,donevaIQRelationsRapidly2015}. Finally, in 
Refs.~\cite{apariciorescoNeutronStarsTheories2016,fernandezRealisticBuchdahlLimit2025}, 
the authors also analyze neutron stars 
in $R-squared$ gravity, however, they take $a<0$, unlike the values $a>0$ used here and in most of the aforementioned 
studies. The resulting NS solutions are not asymptotically flat (AF), contrary to the claims of those authors  (cf. also Ref. \cite{astashenokRealisticModelsRelativistic2017}). In order to 
justify this statement, in Appendix \ref{app:NSanegative} we analyze those type of solutions and show that asymptotically the Ricci scalar 
never reaches the value $R=0$ (unlike the scenario with $a>0$) but rather oscillates around it. This behavior produces 
a mass function $\mu(r)$ that also oscillates and grows asymptotically without reaching a fixed value, the one that would correspond 
to the ADM mass for genuinely AF solutions associated with $a>0$.

 Following a methodology 
 similar to the JPS formalism \cite{jaimeRobustApproachGravity2011}, the analysis in Ref. \cite{capozzielloMassRadiusRelationNeutron2016} studies the mass-radius relationship  by solving directly the full modified  field equations without resorting to the usual STT transformation. Moreover, in addition to the $R-squared$  model, they studied other models like $f(R) = R  + \alpha R^2(1 + \gamma R)$ and $f(R) = R^{1 + \epsilon}$. Those authors used the analytical representations of the so called BSkX EOS \cite{pearsonInnerCrustNeutron2012, potekhinAnalyticalRepresentationsUnified2013a} and the Sly EOS. Within the non-perturbative approach other studies explored different kind of EOS \cite{astashenokFurtherStableNeutron2013,astashenokRealisticModelsRelativistic2017,fengEquationStateNeutron2018,nava-callejasProbingStrongField2023}. The diagram in Fig. \ref{fig:ns_rsquared} 
 (see Appendix \ref{app:summary}) summarizes some of the strategies and methods followed by several authors to tackle NS in $R-squared$ gravity.
 
 Some comments concerning the specific EOS used to construct NS are in order.
 One can use {\it realistic} EOS motivated by microphysics and field theory, which 
 span a whole range of densities, from the inner core to the outer crust of NS. 
 Many of those realistic EOS are given numerically in the form of tabulated values for the 
 pressure, energy-density and baryon density, and therefore, require interpolation before starting the numerical integration of the hydrostatic equilibrium equations, 
whether in GR or in modified theories of gravity.
 
 Another, more economical possibility consists in mimicking those realistic EOS with polytropes that can account for the different layers of NS. The advantage is that 
 the solution is not ``contaminated" by the inherent numerical errors associated with 
 the interpolation method. In this paper we will follow this strategy and use different 
 kind of EOS based on polytropes to model NS within $R-squared$ gravity. 
 We use a non-perturbative approach with a method different from \citet{yazadjievNonperturbativeSelfconsistentModels2014}, but our results are entirely consistent. We then compare our results with the perturbative studies in order to provide and independent assessment about the reliability of the perturbative approach. 

 As we illustrate below,  one of the most interesting features of $f(R)$ gravity is that
 one can obtain NS models with masses that are larger than the maximum mass of NS in GR for exactly the same EOS. Therefore with this kind of theories one can in principle account for the large masses of recently observed NS of about $2.35\pm 0.17 M_\odot$ \cite{romani2022} that cannot be obtained from GR with the same EOS used here.

As we mentioned in the Introduction, 
our goal is to review the analysis of the $R- squared$ model in the context of NS and update previous studies with current observations, and novel techniques. In particular, we use a more sophisticated piecewise polytrope EOS that mimics realistic EOS for a wide range of densities within a neutron star. In fact, we even consider an incompressible (constant-density fluid) inside the star in order
to contrast these results with those of the previous section to show that the problems we have to compute a ``realistic" model for the Sun are not related with the use of this kind of simplified EOS but rather with the different scales involved in the cosmological $f(R)$ models and the scales associated with the Sun. In the case of the $R-squared$ model, the scale involved in the parameter $a$ is precisely of the same order of length scales found in NS, and this is why we do not 
encounter any numerical drawbacks to find ``realistic" NS models 
even with this kind of constant-density matter.

\subsection{The incompressible NS models}
\label{sec:nsrhocons}

In this section we show the numerical results of NS models constructed with constant-density stars for $R-squared$ gravity. Although we know that
this kind of {\it incompressible} matter is not realistic, it
reproduces the bulk properties of a NS with a good approximation, like the mass and radius. In the next section we consider more realistic matter models for NS.

For the specific $R-squared$ model at hand \eqref{eq::f_R_RSquered_model} it is instructive to take a closer look to the equation for the Ricci scalar \eqref{eq::eq::Rpprime_1}:
\begin{equation}
\label{eq::RSquared}
    R''= \frac{1}{6a}\left[m(\kappa T + R)\right] +\left(\frac{m'}{2m}-\frac{n'}{2n}-\frac{2}{r}\right)R'.
\end{equation}
In contrast to the cosmological $f(R)$ models analyzed in the previous section,  the potential $V_{eff}(R,0)= R^2/6 + {\rm const}$ has a global minimum at $R=0$, and consequently, this model cannot be used to explain the late cosmological 
acceleration in a natural way, as the effective cosmological constant vanishes in cosmic time.
Therefore, in the context of stars, the Ricci scalar interpolates from a finite value at the center of the star to a vanishing value asymptotically. This is 
regardless of the EOS since the star will have a compact support and the fluid variables vanish outside the star by definition, and furthermore, because the 
spacetime will be asymptotically flat as $R$ vanishes very rapidly asymptotically. This
property will apply as well when dealing with more realistic EOS where a constant-density NS is no longer assumed (see the next section).

\begin{figure}
\centering
\includegraphics[width=\textwidth]{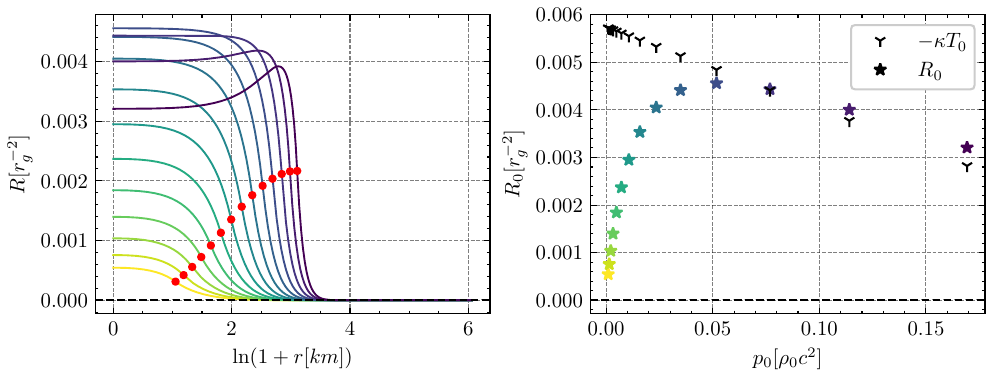}
\caption{{\it Left panel:} Ricci scalar profile for the $R-squared$ gravity model with $a=r_g^2\approx 2.2 {\rm km}^2$, for different values of the central pressure $p_0$. Colors from violet (higher pressure) to green (lower pressure) label the solutions for different $p_0$. The black stars mark the location of the 
surface of the star (i.e. the NS radius) where the pressure vanishes. The dotted horizontal lines (left panel) indicate the values $R=-\kappa T_0$ in GR inside the star, where $T_0=3p_0-\rho c^2$ at $r=0$. {\it Right panel:} Ricci scalar at $r=0$ as a function of the central pressure for the $R-squared$ model (stars) and in GR $R_0=-\kappa T$ (\texttt{Y} black marks). The star-markers colors are associated with the corresponding colors of solutions shown in the left panel.}
\label{fig::RSQuared_Incompressible_Fluid_Ricci_Pressure}
\end{figure}

Let us start by describing the setup. We assume a step function with the value $\rho= 10^{14} \rho_\odot$ for the interior solution, and a null density outside the star, and modify the central pressure to obtain different hydrostatic configurations. For simplicity we also adopt the value $a=r_g^{2}$ ($r_g= GM_\odot/c^2\approx1.48 \text{ km}$, see Appendix \ref{secc::Numerical_Strategy}), since for these simplified NS models it is not worth to perform a thorough examination of the parameter $a$. In the next section we fill this gap for different and more realistic NS models. Figure \ref{fig::RSQuared_Incompressible_Fluid_Ricci_Pressure} (left panel) depicts multiple solutions with decreasing central pressures (colors violet to green) for the Ricci scalar 
$R$. Note that as the pressure increases there is 
a {\it critical value} of $p_0$ beyond which $R$ does not fall off monotonically, but instead, increases and develops a local maximum inside the star, and then decreases monotonically towards its vanish value at spatial infinity. 
This feature is related with the change of sign in $R''$ at $r=0$ according to the 
regularity condition \eqref{Rbiprime0} where $R''(0)=2R_2=[\kappa(3p_0-\rho c^2)+R_0]/18a$. Thus, when $p_0$ and $R_0$ are low  $R''(0)$ is negative and 
dominated by the term with $-\rho$. Then the Ricci scalar decreases monotonically from the center of the star to the asymptotic value $R=0$. However, as $p_0$ and $R_0$ increases, $R''(0)$ becomes positive 
and $R$ starts increasing until it reaches a local maximum and then it decreases again monotonically.
The right panel of Fig.\ref{fig::RSQuared_Incompressible_Fluid_Ricci_Pressure} displays the Ricci scalar at the center of the object as a function of the central pressure. The three-peak black marker (\texttt{Y}) represents the value of the trace of the energy-momentum tensor that belongs to the exact value in GR ($R_{GR}=-\kappa T= -\kappa(3p-\rho c^2)$, computed at $r=0$) that we included for reference. For central pressures $0.05 \rho_0 c^2\lesssim p_0$ we observe a correlation and a close agreement between $R_{GR}$ and $R$ [the latter computed by solving Eq. \eqref{eq::RSquared}, and both evaluated at $r=0$]. However, for lower values of the central pressure the relative difference between both quantities becomes much larger.

\begin{figure}
\centering
\includegraphics[width=\textwidth]{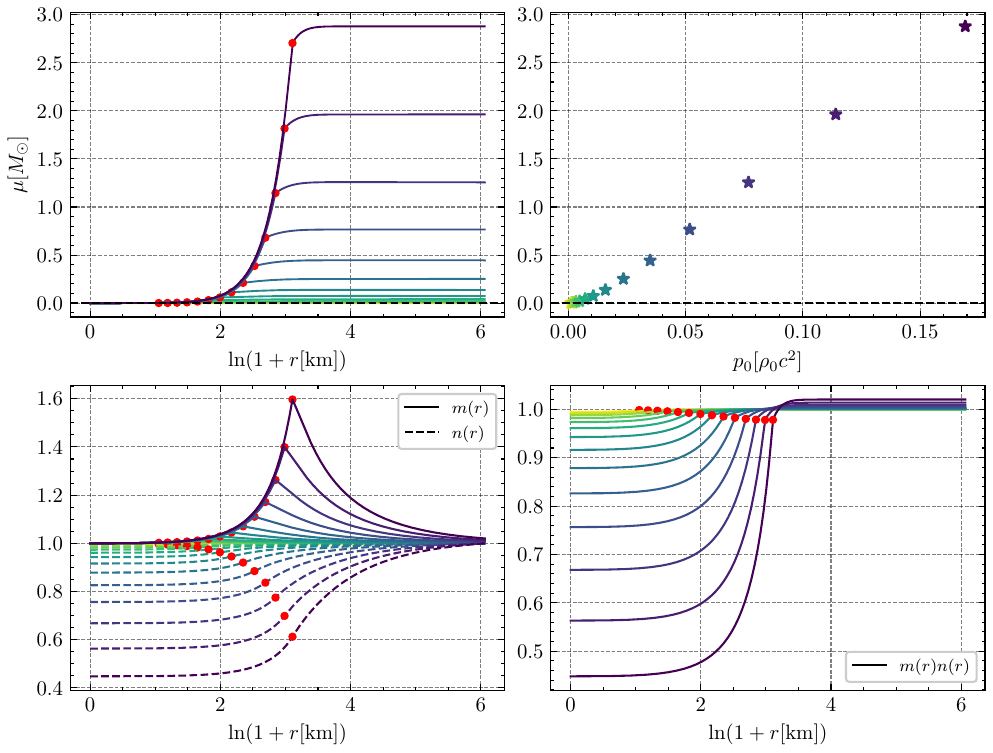}
\caption{{\it Top left panel:} Mass function $\mu(r)$ associated with the solutions of Fig.\ref{fig::RSQuared_Incompressible_Fluid_Ricci_Pressure}. 
The red circles indicate the location of the star radius. The asymptotic value of the mass function corresponds to the total gravitational mass of the NS. {\it Top right panel:} the gravitational mass of the NS
for different values of the central pressure $p_0$ used for 
computing the NS (constant-density) configurations in $R-squared$ gravity. {\it Bottom left panel:} Metric potentials $m(r)$ and $n(r)$. {\it Bottom right panel:} Product of the metric potentials $m(r)\times n(r)$.}  
\label{fig::Multiple_press_Ricci_metric_a_1_mass}
\end{figure}

Fig. \ref{fig::Multiple_press_Ricci_metric_a_1_mass} (top left panel) shows the mass function $\mu(r)$ Eq. \eqref{massfunct} for the same configurations displayed in  Fig.\ref{fig::RSQuared_Incompressible_Fluid_Ricci_Pressure}.
Unlike NS models in $GR$, the mass function still grows outside the star in $R-squared$ gravity due to the contribution of the Ricci scalar itself to the total (effective) energy-density [cf. the r.h.s of Eq. \eqref{eq::fR_formalism_4}].  This feature is also manifested in that $m(r)\times n(r)\neq 1$ outside the star (cf. the bottom right panel). The top right panel illustrates the behavior of the total gravitational mass of the NS, given by $M=\mu(\infty)$, with respect to the central pressure. We observe almost a linear relation between both quantities. This simple model can reproduce the observed mass and radius of NS stars given some central pressure as we can observe from these plots 
and from Table \ref{table::RsquaredIncompressible}.

\begin{table}[!htb]
\begin{center}
\begin{tabular}{||c c c||}
\hline
$p_0(\rho_0c^2)$ & $M[M_\odot]$ & $R_{NS}[\text{km}]$ \\
 \hline\hline
0.16 &  2.87 & 21.377\\
0.11 & 1.96 & 18.81\\
0.076 & 1.25 & 16.22\\
0.051 & 0.765 & 13.74\\
0.034 &0.443 & 11.47 \\
0.023& 0.252& 9.47 \\
0.015& 0.137& 7.76 \\
0.010 & 0.074 & 6.33\\
0.0071 & 0.042&5.16\\
0.0048& 0.020& 4.205\\
0.0032& 0.0140& 3.427\\
0.0022& 0.0039& 2.796\\
0.00148&0.0051&2.2850\\
$10^{-3}$&0.0003& 1.868\\
\hline
\end{tabular}
\end{center}
\caption{Masses (and their correspondent radii $R_{NS}$) for an incompressible fluid in {\it R-squared} gravity obtained by varying the central pressure for the configurations shown in Figs. \ref{fig::RSQuared_Incompressible_Fluid_Ricci_Pressure} and \ref{fig::Multiple_press_Ricci_metric_a_1_mass}. }
\label{table::RsquaredIncompressible}
\end{table}

We close this section and recapitulate by stressing that in this kind of NS incompressible models under $R-squared$ gravity, we do not encounter any numerical obstacles like in Sec. \ref{Sec::Viablef_Rmodels} within $f(R)$ cosmological models, since the scale involved in the parameter $a$ is of the same order of magnitude of the scales found in NS, and therefore there are no high density contrasts between the NS density and the inherent density $\rho_*= 1/(\kappa a c^2)$ associated with this kind of gravity model. In this way we have strong evidence that those kind of numerical complications under the full non-linear treatment are not due to the use of a constant-density fluid within $f(R)$ gravity, but due to the high density contrast of about $29$ orders of magnitude between the cosmological densities involved in those $f(R)$ models and the Sun's density $\rho_\odot$. 
Those kind of numerical challenges were already encountered in a semilinear analysis \cite{negrelliSolarSystemTests2020}, but were circumvented by adapting several approximations in the equations.

\subsection{Realistic NS models: piecewise polytropes}
\label{sec:polytrop}
The EOS of a NS is rather uncertain, specially at or beyond nuclear densities. Strictly speaking, the whole density range of a NS cannot be accurately approximated with a single EOS, let alone with only one polytrope. When using realistic EOS (most of which are given by numerical tables), one usually matches different EOS for the core and the crust of NS. Thus, 
if one is to model accurately the whole range of densities of a NS with an explicit algebraic EOS, like a polytrope, one needs in 
fact several of them glued continuously at different layers. In this section we incorporate such strategy and 
employ a set of piecewise parametrized polytropic (PPP) EOS that mimics more realistically the matter inside a NS. In particular, the core of NS is modeled in this work by mimicking three realistic EOS, the SLy \cite{douchinUnifiedEquationState2001}, the APR4 \cite{akmalEquationStateNucleon1998} and H4 \cite{lackeyObservationalConstraintsHyperons2006} EOS with piecewise polytropes \cite{readConstraintsPhenomenologicallyParameterized2009}. The details of this construction is found in Appendix \ref{sec::Appendix_Piecewise polytrope EoS}.

\begin{figure}
\centering
\includegraphics[width=\textwidth]{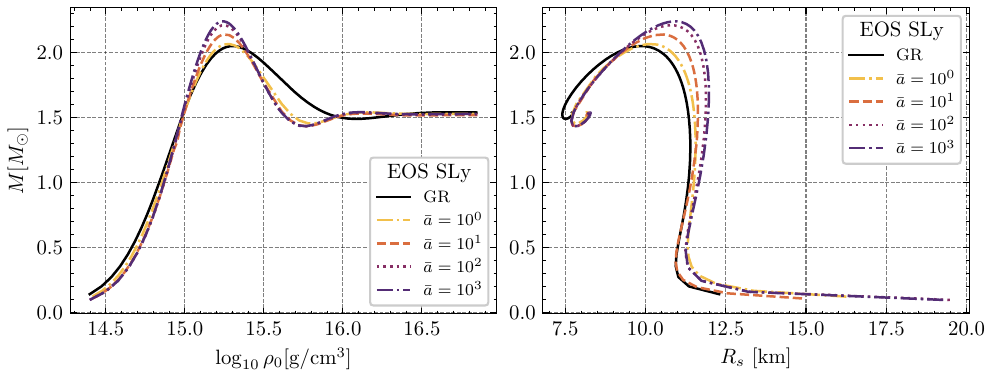}
\caption{The total mass-central-density relation for EOS SLy (left panel) and its mass-radius curve (right panel) for the {\it R-squared model} with three different values of $\bar{a}$. For comparison, the prediction from GR is shown in solid black lines.} 
\label{fig::RsquaredMass_radius_dens_SLY_}
\end{figure}

\begin{figure}
\centering
\includegraphics[width=\textwidth]{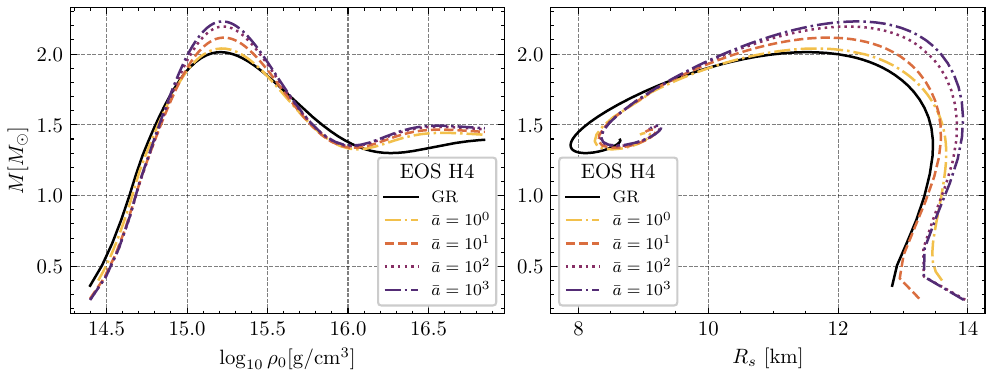}
\caption{Same as Fig.\ref{fig::RsquaredMass_radius_dens_SLY_} but for EOS H4.} 
\label{fig::RsquaredMass_radius_dens_H4}
\end{figure}

\begin{figure}
\centering
\includegraphics[width=\textwidth]{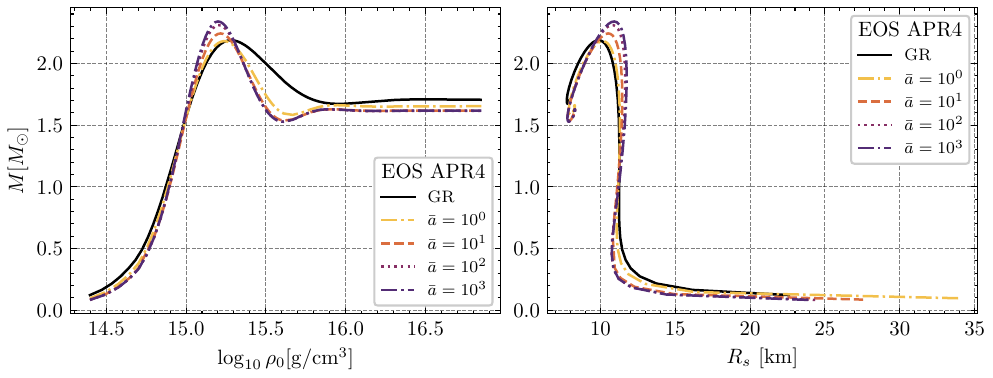}
\caption{Same as Fig. \ref{fig::RsquaredMass_radius_dens_SLY_} but for EOS APR4.}  
\label{fig::RsquaredMass_radius_dens_APR4}
\end{figure}

We generate mass-density and mass-radius relations for different values of the parameter $a$, from ${\bar a}=0$ (GR limit) to ${\bar a}=10^3$ (where ${\bar a}=a/r_g^2$, as in the previous section). Results are depicted in Figures \ref{fig::RsquaredMass_radius_dens_SLY_}--\ref{fig::RsquaredMass_radius_dens_APR4}. 
In the GR case, the three EOS provide maximum mass models with $M_{\max} \lesssim 2.19 M_\odot  $
(see  the second column of Table \ref{table:PPP_Maximal_Masses_with_asq}), in particular, the APR4 EOS provides the largest maximum-mass model $M_{\max}\approx 2.19 M_\odot$, which is still below the mass $\sim 2.35 M_\odot$ observed recently in the rapidly rotating pulsar PSRJ0952-0607 \cite{romani2022}. However, for all the three EOS, the $R-squared$ model with ${\bar a}=10^3$ (sixth column of Table \ref{table:PPP_Maximal_Masses_with_asq})
produces maximum mass models $2.23 M_\odot \lesssim M_{\max}$, which agree at $1\sigma$ or better with the large observed mass in PSRJ0952-0607, in particular, this is the case for the EOS APR4 where $M_{\max}\approx 2.34 M_\odot$.

From Figs. \ref{fig::RsquaredMass_radius_dens_SLY_}--\ref{fig::RsquaredMass_radius_dens_APR4} it is also apparent that the smaller the value ${\bar a}$ the smaller the  differences of $R-squared$ gravity with respect to GR in NS models lying in the stable branches. Even for ${\bar a}\sim 1$ there 
are no significant deviations relative to GR, and in the GR limit  ${\bar a}\ll 1$, one expects (according with the arguments given at the beginning of Sec. \ref{sec:starsR2}) that an $R-squared$ model with such a low value of ${\bar a}$ would not produce 
significant deviations relative to GR in any extended object with scales larger than kilometers, like any self-gravitating object observed so far in the universe such as neutron stars, white dwarfs and ordinary stars.

\begin{figure}
    \centering
    \includegraphics[width=1\linewidth]{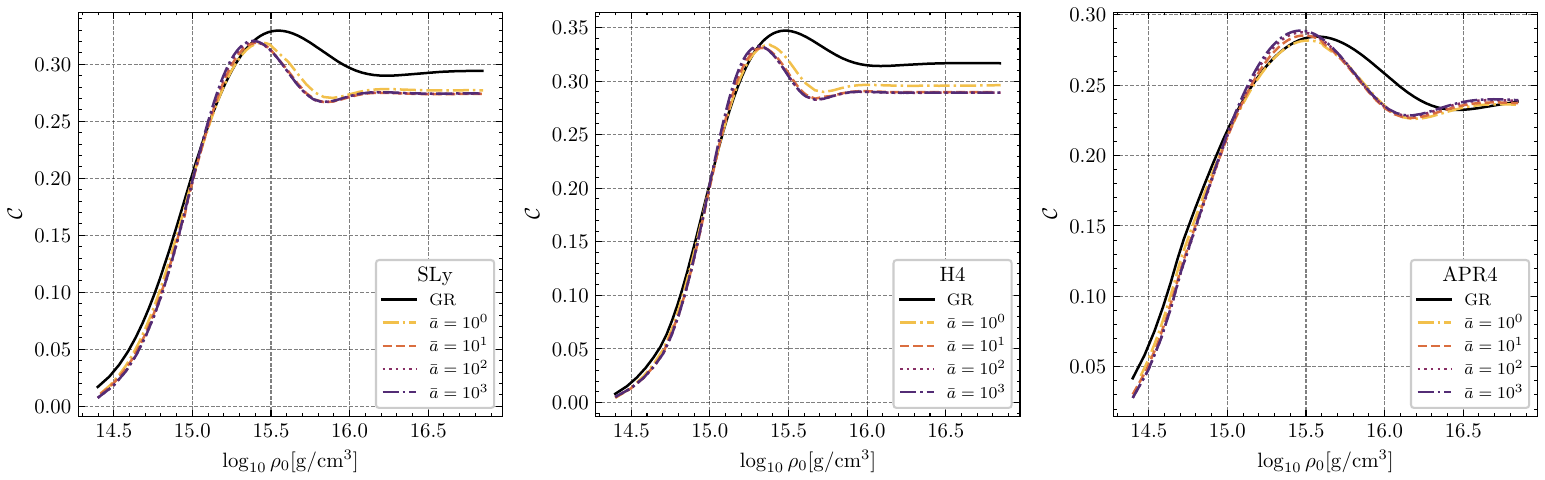}
    \caption{Neutron-star {\it compactness} $\mathcal{C}= \frac{GM}{c^2 R_s}$, defined in terms of the 
    total mass $M$ and radius $R_s$, as a function of the central density for the EOS SLy (left panel), H4 (middle panel) and APR (right panel), 
    and for different values of the parameter $\bar{a}$.}
    \label{fig::Compactness_threepanelEOS}
\end{figure}

Figure \ref{fig::Compactness_threepanelEOS} shows the {\it compactness} 
${\cal C}= GM/(c^2R_s)$ associated with each NS model. For the stable branches $(M< M_{\rm max})$ 
the compactness in GR (${\bar a}=0$) and 
{\it R-squared} gravity is very similar. 
In GR the compactness turns out to be slightly 
higer than 
in {\it R-squared} gravity 
for the SLy and H4 EOS as the maximum mass 
models are concerned. However, for the APR4 EOS it is the opposite. But again, these differences 
are less than $5\%$ in the stable branches.

\begin{figure}
\centering
\includegraphics[width=\textwidth]{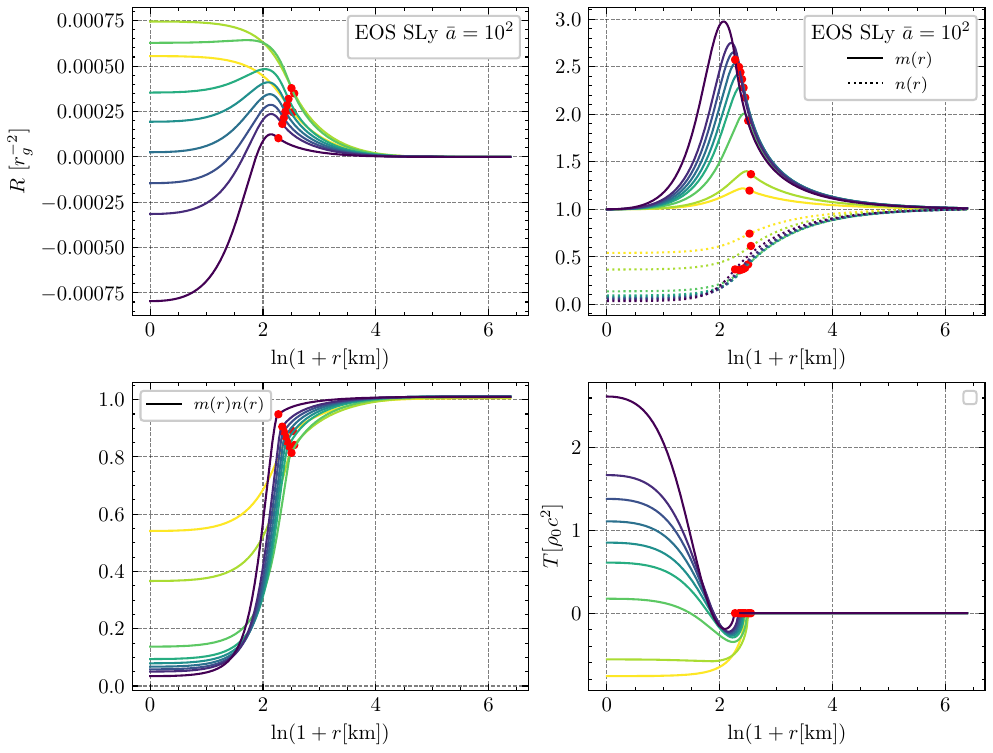}
\caption{{\it Top left:} Ricci scalar solutions for the $R-squared$ model with $a= 100 r_g^{2}\sim 220 {\rm km}^2$ for various central densities $\rho_0$ using the SLY EOS (colors from yellow to violet indicate 
increasing $\rho_0$ and $p_0$). The Ricci scalar vanishes asymptotically for all the solutions. {\it Top right:} metric functions $m$ (solid line) 
and $n$ (dotted lines). {\it Bottom left:} Product of the metric functions $n(r)\times m(r)$. 
Unlike GR, this product is not exactly one beyond the star's surface  since the 
solution exterior to the star is not the Schwarzschild solution. However, the solution is asymptotically flat. {\it Bottom right:} Behavior of the trace $T=3p-\rho c^2$ with respect to the radial coordinate. Notice that as
$p_0$ increases the star becomes more relativistic (the metric function $n_0$ decreases) and $T_0$ changes 
sign. In all the plots the red dots indicate the location of the star's surface 
(i.e. the star's radius) where the pressure and density vanish.}
\label{fig::Multiple_dens_Ricci_metric_SLY_100}
\end{figure}

Figure \ref{fig::Multiple_dens_Ricci_metric_SLY_100} shows a sample of solutions for the Ricci scalar $R(r)$ (top left panel) and the two metric functions $n(r)$ and $m(r)$ (top right panel) with respect to the coordinate $r$ for multiple central densities associated with the SLy EOS. Each point of Fig. \ref{fig::RsquaredMass_radius_dens_SLY_} represents a NS configuration associated with one of these solutions obtained after performing a shooting method such that $R(r)\rightarrow 0$ as $r\rightarrow \infty$. We observe that 
the metric functions approach the Minkowski values asymptotically ($n\rightarrow 1,m\rightarrow 1$) . Moreover, 
as in the case of the constant-density NS, the product of the 
metric functions $n(r)\times m(r)\neq 1$ (bottom left panel)
outside the star where the fluid's pressure and density vanishes. This is because the 
Ricci scalar contributes to the total effective energy-density, and both quantities do not vanish 
outside the star as one can see from the top left panel of this figure. In other words, 
the Birkhoff theorem does not apply in general for $R-squared$ gravity, and thus, outside 
the star the spacetime does not corresponds to the Schwarzschild solution. The color code in the plots indicate lower (yellow color) to higher densities (violet color), 
and the Ricci scalar is presumably \textit{unscreened}, in the sense that screened solutions usually show a Ricci scalar that would follow closely the GR value $-\kappa T$ inside and outside the star. This means that in screened solutions, $R$ usually vanishes or decreases very rapidly but continuously outside the surface of the star. 
As we increase the central density $\rho_0$, and so the pressure 
$p_0$ via the EOS, we notice the emergence of solutions with negative curvature at the center of the NS. This is related to the fact that $T_0=3p_0-\rho_0 c^2$ is negative for low pressures $p_0$, but for sufficiently large $p_0$, $T_0$ becomes positive (cf. the bottom right panel of 
Fig. \ref{fig::Multiple_dens_Ricci_metric_SLY_100}) and this 
quantity appears in the differential equation for $R$. Thus, the sign of 
$R''$ at $r=0$ is directly correlated with the sign of 
$T_0$ [cf. Eq. \eqref{eq::RSquared}]: $R''$ is negative ($R$ positive) for low pressures and positive ($R$ negative) for larger ones. As stressed before, in GR $R\equiv -\kappa T$ exactly, and the
correlation between the alluded signs is more evident, leading to a vanishing $R$ immediately outside the star. In $R-squared$ gravity this correlation is approximate and $R$ does not 
vanish in the neighborhood of the star's surface, as it is apparent from the plots.

For the H4 and APR4 EOS the Ricci scalar profiles and metric functions exhibit the same qualitative behavior as those shown Fig. \ref{fig::Multiple_dens_Ricci_metric_SLY_100}.

\begin{table}[!htb]
\centering
\begin{tabular}{||lccccc||}
\hline
\textbf{EOS} & \boldmath{$\bar{a}=0$ (GR)} 
& \boldmath{$\bar{a}= 10^0$}
& \boldmath{$\bar{a} = 10^1$}
& \boldmath{$\bar{a} = 10^2$}
& \boldmath{$\bar{a} = 10^3$}\\
\hline
SLy  & 2.047 & 2.0634 & 2.1366 & 2.2044 & 2.237 \\
H4   & 2.013 & 2.0392 & 2.115 & 2.194 &2.232  \\
APR4 & 2.187 & 2.179 &  2.242&  2.310& 2.339 \\
\hline
\end{tabular}
\caption{Maximum masses $M_{\rm max}$ (in $M_\odot$) in GR ($\bar{a}=0$ second column) 
and in $R-squared$ gravity for various values of $\bar{a}=a/r_g^2$ (third to sixth columns) and for the three EOS used to construct NS.}
\label{table:PPP_Maximal_Masses_with_asq}
\end{table}

\begin{figure}
\centering
\includegraphics[width=\textwidth]{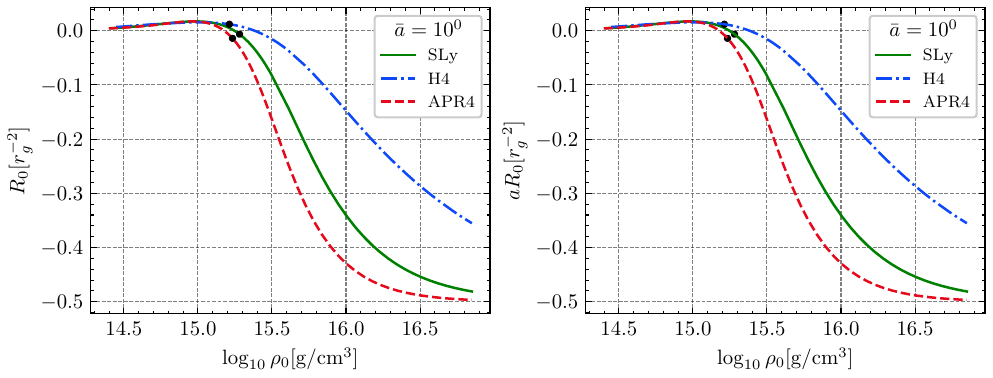}
\caption{{\it Left panel:} Ricci scalar at $r=0$ in units of $r_g^{-2}$ for different values of the central 
density for EOS SLy, H4 and APR4 with $\bar{a}=1$. {\it Right panel:} Same as the left panel but rescaled as $aR_0$. 
In this case both plots are exactly the same.  We observe that $|aR_0|\ll 1$ in the stable branches of NS. Black dots indicate the maximum mass solution. } 
\label{fig::R2ovRa1}
\end{figure}

\begin{figure}
\centering
\includegraphics[width=\textwidth]{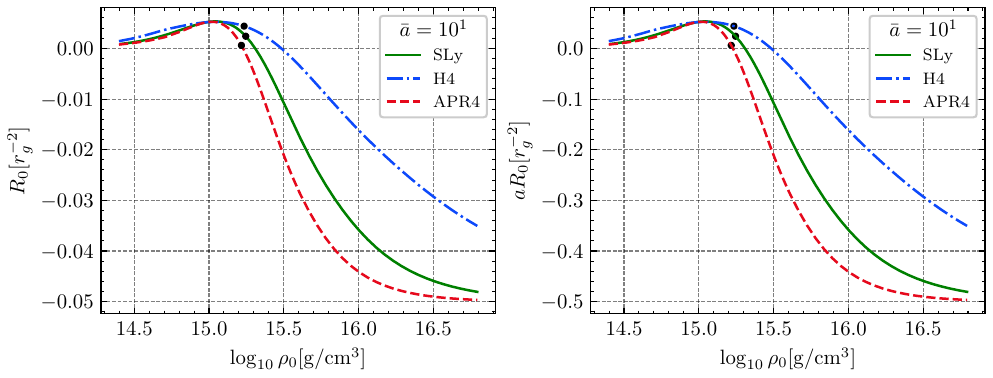}
\caption{{\it Left panel:} Ricci scalar at $r=0$ in units of $r_g^{-2}$  for different values of the central 
density for EOS H4, SLY and APR4 with $\bar{a}=10^1$. {\it Right panel:} Same as the left panel but rescaled as $aR_0$. 
We observe that $|aR_0|\ll 1$ in the stable branches of NS.  Black dots indicate the maximum mass solution. } 
\label{fig::R2ovRa10}
\end{figure}

\begin{figure}
\centering
\includegraphics[width=\textwidth]{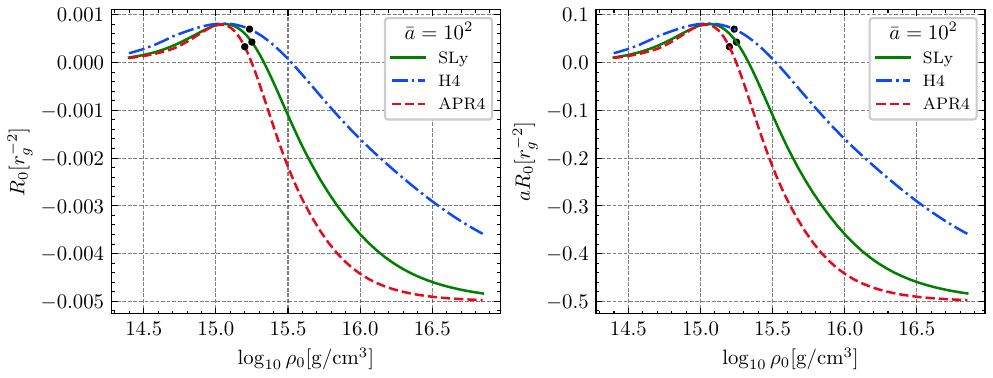}
\caption{{\it Left panel:} Ricci scalar at $r=0$ in units of $r_g^{-2}$  for different values of the central 
density for EOS H4, SLY and APR4 with $\bar{a}=10^2$. {\it Right panel:} Same as the left panel but rescaled as $aR_0$. 
We observe that $|aR_0|\ll 1$ in the stable branches of NS.  Black dots indicate the maximum mass solution. } 
\label{fig::R2ovRa100}
\end{figure}

\begin{figure}
\centering
\includegraphics[width=\textwidth]{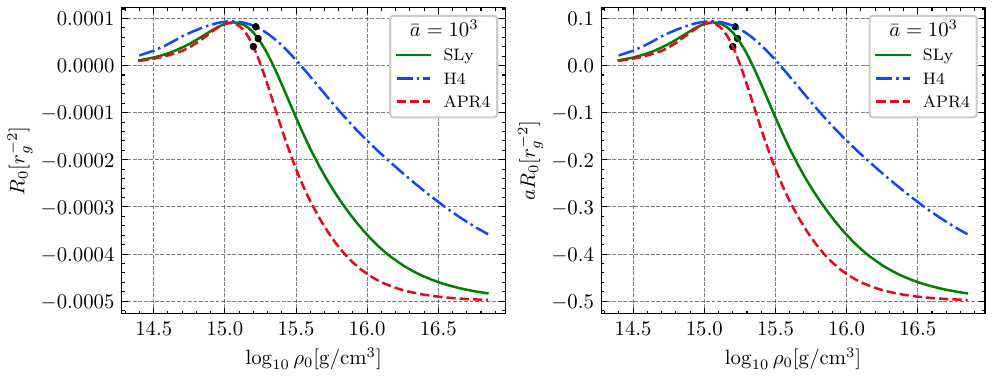}
\caption{{\it Left panel:} Ricci scalar at $r=0$ in units of $r_g^{-2}$ for different values of the central 
density for EOS H4, SLY and APR4 with $\bar{a}=10^3$. {\it Right panel:} Same as the left panel but rescaled as $aR_0$. 
We observe that $|aR_0|\ll 1$ in the stable branches of NS.  Black dots indicate the maximum mass solution. } 
\label{fig::R2ovRa1000}
\end{figure}

\begin{figure}
\centering
\includegraphics[width=\textwidth]{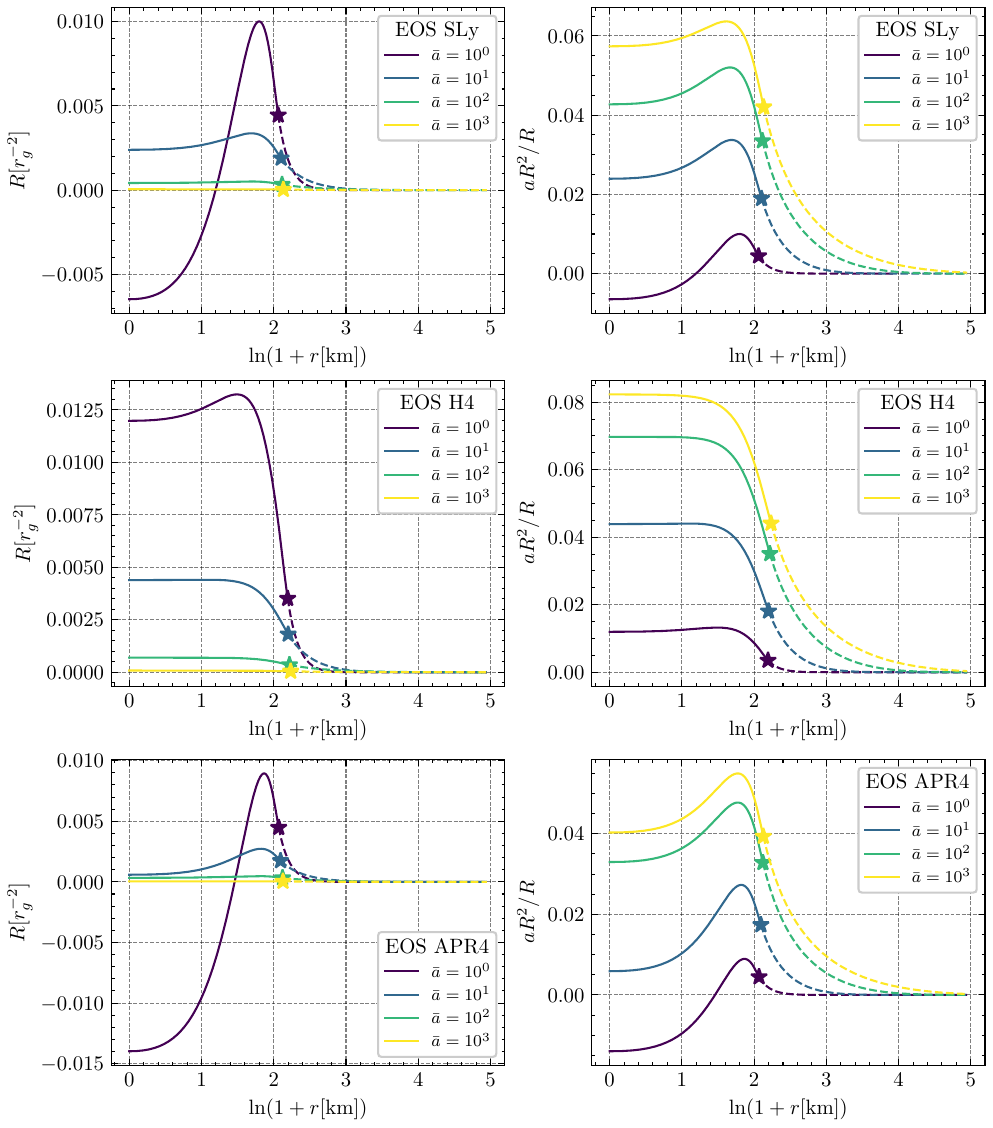}
\caption{{\it Left columns:} Ricci scalar 
associated with the maximal mass configuration for each EOS SLy, H4 and APR4, for multiple values of $\bar{a}$.
{\it Right columns:} Dimensionless contribution 
$aR$ to $R-squared$ gravity with $R$ given by the respective left panel. Notice that such contribution 
is relatively small compared with unity inside and outside the NS. The color stars indicate the location of the 
NS star radius.} 
\label{fig::RaH4}
\end{figure}

Figures \ref{fig::R2ovRa1}--\ref{fig::R2ovRa1000} show the Ricci scalar at the center of the NS for the three EOS 
H4, SLY and APR4 for different values of the parameter $a$ and  for all the central densities considered. 
We conclude that $|aR_0|\ll 1$ holds for central densities in the stable branch of NS and that this feature is satisfied for $aR(r)$ at all the places inside and outside the NS, not only at $r=0$ as one can see from Fig. \ref{fig::RaH4}. This result reveals that, in principle, a perturbative approach 
based on the assumption $aR\ll 1$ would be a good approximation. However,  it could also happen that the small errors associated with terms of higher order 
in $aR$ can accumulate during the numerical integration in a way that the global quantities like the gravitational mass of the NS, might be quite different relative to the non-perturbative treatment. In fact, and as already remarked in Ref.\cite{yazadjievNonperturbativeSelfconsistentModels2014}, the perturbative approaches like in Refs.\cite{cooneyNeutronStarsGravity2010,arapogluConstraintsPerturbativeGravity2011} tend to under estimate the NS masses (in particular the maximum mass) for $a>0$, which is the only sector that is physically relevant and which 
is associated with the positivity of $f_{RR}=2a$ (cf. Appendix \ref{app:NSanegative} for the sector $a<0$). In our case, the NS masses tend to grow as $a$ increases. However, this increase in the mass is more noticeable for ${\rm km}^2 \ll a$, but for 
$10^4 {\rm km}^2 < a$ the maximum mass and radius reach a limit as $a$ increases which is 
close to those reported in Table \ref{table:PPP_Maximal_Masses_with_asq} for $a=10^3 r_g^2$. 
This trend was also reported in Refs.\cite{yazadjievNonperturbativeSelfconsistentModels2014,nava-callejasProbingStrongField2023}. For $a\lesssim {\rm km}^2 $, the differences relative to GR are almost negligible and can also be explained by slight changes in the EOS as acknowledged in Ref. \cite{cooneyNeutronStarsGravity2010}. Therefore, $R-squared$ gravity for small values of $a$ is neither constrained by NS nor predicts new and important effects, at least as the bulk properties of NS are concerned. On the other hand, our non-perturbative analysis agrees entirely with the results found in Refs. \cite{yazadjievNonperturbativeSelfconsistentModels2014,nava-callejasProbingStrongField2023}, and confirms that the
robust approach developed in \cite{jaimeRobustApproachGravity2011} makes unnecessary the use of the back-and-forth 
transformations from the Einstein to the Jordan frames and vice versa,  and therefore, it avoids the potential pitfalls when the back-and-forth transformations are not performed 
self-consistently, namely, when reporting physical observables. But most importantly, the full non-perturbative analysis does lead to the values $a$ for which the $R-squared$ model is able to reproduce the observed mass of PSRJ0952-0607 of $\sim 2.35 M_\odot$ without the need of stiffer or exotic EOS. 
Needless to say, this kind of modified gravity models that affect to a large extent the properties of NS should face also the constraints imposed by binary pulsars, in particular, those related with the emission of gravitational 
waves (GW). This is because $f(R)$ gravity, like STT, can propagate a scalar degree of freedom (SDOF) in the form of scalar GW. Therefore, if that kind of radiation is sufficiently large, it can spoil the observed dynamics of such binary systems. But, if the SDOF turns to be small or suppressed, then its effects on a NS can be 
negligible, and then $R-squared$ gravity would not explain by its own the high mass of PSRJ0952-0607. Finally, and related to these issues,  the values for the parameter $a$ of the $R-squared$ model cannot be {\it extremely} high, as otherwise the theory might also spoil the solar system tests 
(if $a> 5\times 10^5 {\rm km}^2$) or the double binary pulsar PSR J0737-30-39 precession rate 
(if $a> 2.3 \times 10^9 {\rm km}^2$) \cite{NafJetzer2010}.

\section{Conclusions}
\label{sec: conclusion}
We have analyzed $f(R)$ gravity in two scenarios that involves two completely 
different regimes and models. In the first case, we studied some 
cosmologically-motivated $f(R)$ models that were proposed in the past to mimic 
dark energy, but in order to check if they pass or not the solar system tests while
analyzing the full set of non-linear equations in a non-perturbative fashion. 
We concluded that the large difference of scales between the low built-in cosmological density involved in such $f(R)$ models and the relatively high solar density (appearing in the fluid variables describing the Sun) makes the problem to be intractable numerically even with high accurate numerical schemes. Such a high contrast in densities that involves 29 orders of magnitude  proved to be a formidable numerical obstacle, as far as number-crunching algorithms are concerned. In the past, these obstacles have been overcome by using approximations in the equations through perturbative approaches that lead to linear or semi-linear equations, followed by the use of software that can handle more easily arbitrary numerical precision (only limited by memory allocation) like {\it MATHEMATICA} \cite{negrelliSolarSystemTests2020}. It remains to be investigated if we can implement this kind of software to treat the full non-linear set of equations and corroborate if the chameleon-like mechanism or {\it screening} that suppresses the scalar DOF outside the Sun emerges naturally from the equations or not, and thus, if 
such $f(R)$ models lead or not to a PPN parameter $\gamma \sim 1$ within the solar system.

In the second scenario we analyzed the so-called $R-squared$ model by adjusting 
the parameter $a$ of the quadratic term $aR^2$ in order to test neutron-star properties.
We performed such study also non-perturbatively and still working with a formulation in the Jordan frame that avoids the use of conformal transformations and the inversion of variables that involve the introduction of a scalar-field $\phi= f_R$. By doing so we deal only with second-order equations at most, namely, a second-order equation for the Ricci scalar itself. We also adapted several polytropic EOS that mimic realistic EOS describing the different layers of a NS, and compared our results with the recently observed heaviest NS detected to date of $\sim 2.35 M_\odot$ \cite{romani2022}. We showed that GR cannot safely reproduce this mass with the same kind of EOS and in the non-rotating
limit. However, $R-squared$ gravity can satisfy this observational constraint provided 
the parameter $a$ is as high as $\sim 2190 {\rm km}^2$. Larger values for $a$ 
do not lead to significant larger NS masses, but may otherwise suffer from observational constraints at the level of the solar-system tests \cite{NafJetzer2010}. On the other hand, for lower values, 
for instance, $a\sim {\rm km}^2$, the difference with GR is small, but sufficiently high to reveal that the perturbative approaches tend to underestimate the NS masses. Moreover, these differences are small enough as to be masked by the 
EOS themselves, and therefore, such low values of $a$ in $R-squared$ gravity cannot be really tested, at least as concerns the bulk properties of NS. 

There are other scenarios where $R-squared$ gravity might face stringent bounds with 
$ a \gg {\rm km}^2  $ that are perhaps worth analyzing in the future. For instance, binary pulsars can constrain those large values for $a$ because the models might predict a higher emission of gravitational radiation in the form of scalar GW in addition to the usual tensor (two-polarization mode) 
GW predicted by GR. Finally, since this gravity model does not produce a viable late cosmological accelerated expansion, it still requires a cosmological constant or other non-quadratic $R-$terms with a 
built-in dark-energy scale to do so. Future GW detections, cosmological observations and heavier NS discoveries would shed light about the need or not of a modified theory of gravity like the one encoded in $f(R)$ theory or in similar proposals that include torsion\cite{ghoshMaximumMassLimita}, and other geometric scalars coupled or not to fundamental scalar-fields \cite{gaussbonet}.


\section{Acknowledgments}

This work was partially supported by 
DGAPA-UNAM grant IN105223 and Conacyt grant 140630. M.S. acknowledges support from DGAPA-PASPA sabbatical grant.

\clearpage

\appendix
\section{Dimensionless form of Stellar Structure Equations for $f(R)$ gravity}
\label{secc::Numerical_Strategy}
The numerical strategy chosen to solve the modified TOV equations in $f(R)$ gravity  is based on a higher-order adaptive size Runge-Kutta (RK) method that implements a shooting technique. This is a widely used technique to solve boundary value problems for differential equations, which we discuss in more detail below. In order to implement the method, the first step is to write the set of differential equations of Section \ref{sec::f_R_Static_Spheerical} in a suitable dimensionless form, starting with the following variables: 
\begin{align}
\label{eq::F_R_Adim_quantities}
    r & = \hat{r} r_\star,& R = \hat{R}R_\star,&  &   \rho & = \hat{\rho}\rho_\star, & p&=\hat{p}p_\star,& \hat{f}_{R_l}& = R_\star^{l-1} f_{R_l},
\end{align}
where the  index $l$ indicates the order of the derivative respect $R$. In general, the models we are considering can be seen as:
\begin{equation}
\label{eq::F_R_dimensionless}
    \hat{f}(\hat{R}):=\frac{f(\hat{R})}{R_\star} = \hat{R}+ \eta F_1\left(\frac{\hat{R}}{\eta}\right),
\end{equation}
where $\eta := R_*/R_\star$ is the fraction between a characteristic scale proportional to the cosmological scale $R_*$ and the unit $R_\star$ selected to measure $R$. Since we are considering a perfect fluid, it follows from  Eq. \eqref{eq::F_R_Adim_quantities} that the dimensionless trace of the energy-momentum tensor is given by
\begin{equation}
    \hat{T} = \frac{T}{p_\star} = \left(-\frac{\hat{\rho}}{b} + 3 \hat{p}\right),
\end{equation}
where, as in GR, $b = p_\star/\rho_\star c^2$. Note that if we choose to measure the pressure in energy density units, then $b=1$. Therefore, using the above dimensionless quantities \eqref{eq::F_R_Adim_quantities}, the set of equations \eqref{eq::eq::Rpprime_1}-\eqref{eq::F_R_TOV_Equations} is recast in the following dimensionless form: 
\begin{subequations}
\label{eq::f_R_TOV_ADIM_EQUATIONS}
\begin{align}
\label{eq::f_R_TOV_ADIM_EQUATIONS_1}
      \hat{R}''&= \frac{1}{3\hat{f}_{RR}}\left[m(\alpha \hat{T}+ \beta(2\hat{f}-\hat{R}\hat{f}_R))-3\hat{f}_{RRR}\hat{R}'^2\right] + \left(\frac{m'}{2m}-\frac{n'}{2n}-\frac{2}{\hat{r}}\right)\hat{R}',\\
       m'= & \frac{m}{\hat{r}(2\hat{f}_{R} + \hat{r}\hat{R}'\hat{f}_{RR})} \bigg\lbrace 2\hat{f}_{R} (1-m)- 2 \alpha m \hat{r}^2 \hat{T}^t_t  + \frac{m\hat{r}^2}{3} \left[\beta (\hat{R}\hat{f}_{R}+ \hat{f}) + \alpha \hat{T}\right]\\\notag
     & + \frac{\hat{r}\hat{R}'\hat{f}_{RR}}{\hat{f}_{R}}\left[\frac{m\hat{r}^2}{3}[\beta (2R\hat{f}_{R}-\hat{f})+ 2\alpha \hat{T}] - \alpha m \hat{r}^2(\hat{T}^t_t+\hat{T}^r_r)+2(1-m)\hat{f}_{R}+ 2\hat{r}\hat{R}'\hat{f}_{RR}\right]\bigg\rbrace, \\
      \label{eq:nprime_1}
     n' =& \frac{n}{\hat{r}(2\hat{f}_{R} + \hat{r}\hat{R}'\hat{f}_{RR})} \left[m\hat{r}^2[\beta(\hat{f}-\hat{R}\hat{f}_{R})+ 2\alpha \hat{T}^r_r]+ 2\hat{f}_{R}(m-1)- 4 \hat{r}\hat{R}'\hat{f}_{RR}\right],\\
    \label{eq:nnprime_1}
      n'' =& \frac{2nm}{\hat{f}_{R}}\left[\alpha \hat{R}^\theta_\theta -\frac{1}{6}[\beta (\hat{R}\hat{f}_{R}+ \hat{f}) +2\alpha \hat{T}]+\frac{\hat{R}'\hat{f}_{RR}}{m}\right] + \frac{n}{2\hat{r}}\left[2\left(\frac{m'}{m}-\frac{n'}{n}\right)+ \frac{n'\hat{r}}{n}\left(\frac{m'}{m}+\frac{n'}{n}\right)\right].
\end{align}
\end{subequations}
The two dimensionless parameters, $\alpha$ and $\beta$, that appear in the above equations have been defined as:
\begin{align}
\label{eq::Numerical_coeficients}
    \alpha &= \frac{8\pi G}{c^4}r_\star^2p_\star, &  \beta = & R_\star r_\star^2,
\end{align}
which together with $b$ and $\eta$ encode the chosen units.  For the cosmological $f(R)$ models presented in Section \ref{Sec::Viablef_Rmodels} we set $R_\star=H_0/c^2$ and  $r_\star= (R_*)^{-1/2}$, so that $\beta=1$. Finally, taking $\rho_\star=\rho_*$ (Eq. \eqref{eq::rho_ast}) implies $\alpha=1$. 

Meanwhile, for the $R-squared$ model of Sec. \ref{sec:starsR2} we select units such that
\begin{align}
\label{eq::RSQUARED_UNITS}
    r_\star &= r_g, & \rho_\star=&M_\odot/r_g^3, & p_\star&= M_\odot c^2/r_g^3, & R_\star&= r_g^{-2},
\end{align}
where we set $r_g = GM_\odot /c^2\approx 1.47473$ km, that is the Sun's half-Schwarzschild radius.  In these units, the coefficients \eqref{eq::Numerical_coeficients} become $\alpha = 8\pi$ and $\beta =1$ while the parameter $a$ is measured in units of $r_g^2$. This is a common and suitable choice for this kind of $f(R)$ theories (see e.g. \cite{yazadjievNonperturbativeSelfconsistentModels2014, capozzielloMassRadiusRelationNeutron2016}).

A crucial aspect to obtain numerical solutions with the desired asymptotic behavior concerns the value of the Ricci scalar  at the center $\hat{R}_0:=\hat{R}(0)$. In the classical version of the shooting method the boundaries are fixed, and thus, a root searching algorithm can be used to find the correct \textit{initial} condition at the center. However, in our case one boundary remains strictly at infinity, thus, for some initial guess of $\hat{R}_0$ slightly smaller or slightly  larger than the right one,  the function $\hat{R}(\hat{r})$ at some distance $\hat{r}>\hat{r}_{\text{crit}}$ will eventually go to $-\infty$ or $+\infty$. We thus determine the correct value of $\hat{R}_0$ up to some distance $\hat{r}_{\text{crit}}$ limited by the numerical accuracy such that $R$ approaches $\hat{R}_1$ are $\hat{r}$ goes to infinity.  

One important feature that needs to be clarified is that the numerical integration proceeds in two stages. Given a central density $\hat{\rho}_0$ and the central pressure $\hat{p}_0$ together with the Ricci scalar $\hat{R}_0$, we integrate the system \eqref{eq::f_R_TOV_ADIM_EQUATIONS} numerically from the center $\hat{r}=0$ to the surface of the star 
$\hat{R}_s$, defined through the pressure as 
 $\hat{p}(\hat{R}_s)= 10^{-40}$. Then, imposing continuity for all the functions at the stellar surface, we integrate outwards the same system but we set $\hat{\rho}=\hat{p}=0$ up to  $\hat{r}_{\text{crit}}$.

The code was implemented in \texttt{JULIA} using the BigFloat package library for arbitrary-precision floating-point arithmetic based on GMPY2, which supports integer and rational arithmetic via the GMP library  and real and complex arithmetic by the MPFR  and MPC  libraries \citet{baileyHighprecisionComputationMathematical2012}. We also employ the DIFFEQ package \cite{rackauckasDifferentialEquationsJlPerformant2017} with a variety of solvers such as the KENCARP58  algorithm.


\section{The Sun scenario in GR (constant density stars and polytropes)}
\label{app:sunGR}
\begin{figure}
\centering
\includegraphics[width=\textwidth]{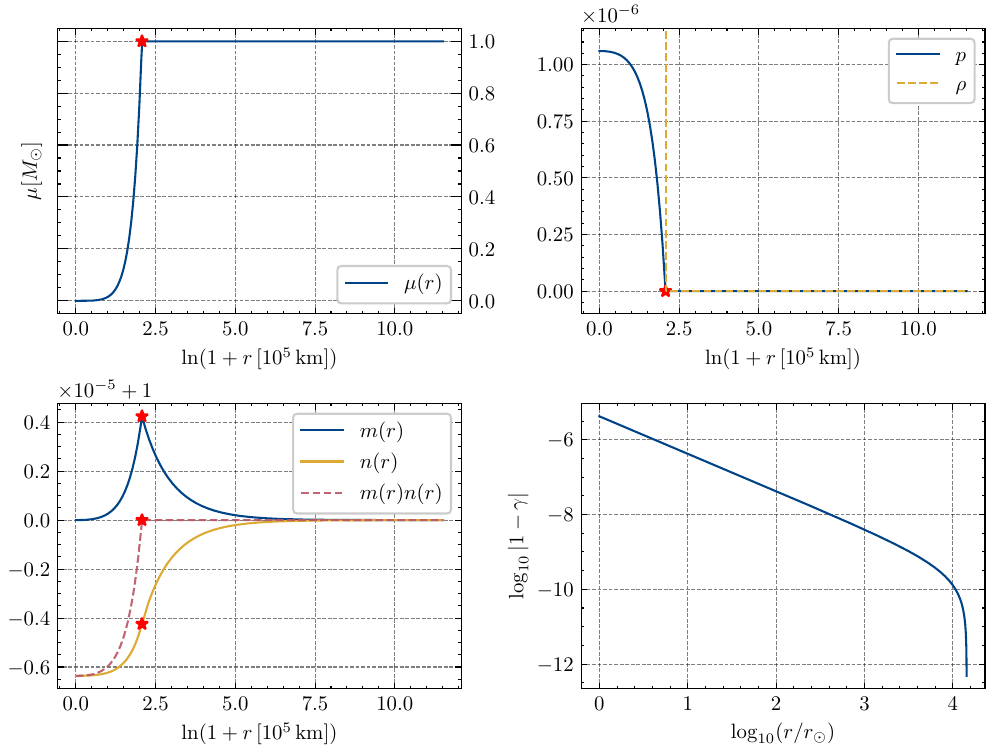}
\caption{
{\it Top left:} Mass function $\mu(r)$ (in units of $M_\odot$) for an incompressible fluid that mimics the Sun. The radial coordinate is in units of  $10^5$ km. The red star indicates the location of the Sun radius. {\it Top right:} Pressure $p(r)$ (solid blue line) in units of the Sun's density $\rho_\odot c^2$ where $\rho_\odot= 1408$ kg/m$^3$ (the yellow dashed line indicates the location of the Sun's surface where the density vanishes; notice the scale, the density is about six orders of magnitude higher) associated with the solution of top-left panel.
{\it Bottom left:} Metric functions  $m(r)$ (solid blue line) and $n(r)$ (solid yellow  line) and 
the product $m(r)\times n(r)$, associated with the solutions of previous 
panels. This product shows that outside the Sun the Schwarzschild solution in this 
coordinates is reached: $m(r)\times n(r)\equiv 1$. The scale of these plots indicates that the deviations with respect to the Minkowski spacetime (i.e. the weak field limit) are very small.  
{\it Bottom right:}  $\gamma$ parameter as function of the radial coordinate as defined in Eq \eqref{eq:GAmma_param_GR_1} outside the Sun. }
\label{fig::SUN_Metric_Gamma}
\end{figure}

One can model the bulk properties of the Sun, like its mass and radius, in a simple 
way by using an incompressible fluid or a polytropic EOS. In the former case, we know 
the exact solution for a constant-density star in GR, and its weak field approximation.
Notwithstanding, it is also useful to have a numerical integrator for that solution. 
In this case, we can take the GR limit without $\Lambda$  ($f(R)=R$) to reduce the system of Eqs. \eqref{eq::F_R_TOV_Equations}:
\begin{subequations}
\label{eq::TOV_m_and_n_variables}
\begin{align}
\label{eq::TOV_m_and_n_variables_m_equation}
\frac{dm}{dr} & = \frac{m}{r}\left[(1-m) + mr^2\kappa\rho\right],\\
\label{eq::TOV_m_and_n_variables_n_equation}
    \frac{dn}{dr} & = \frac{n}{r}\left[mr^2\kappa p +(m-1)\right],\\
    \label{eq::TOV_m_and_n_variables_p_equation}
    \frac{dp}{dr} & = -(p+\rho c^2)\frac{1}{2n}\frac{dn}{dr},
\end{align}
\end{subequations}
which are the TOV equations in GR \cite{waldGeneralRelativity1984}. The system \eqref{eq::TOV_m_and_n_variables_m_equation} is solved numerically for a star with density $\rho=\rho_\odot =1408$ kg/m$^3$ and central pressure fixed by $p_0/\rho_0 c^2\approx 1.06028\times10^{-6}$.  \footnote{The central pressure is fixed through the analytical integration of the pressure function \cite{waldGeneralRelativity1984} 
\begin{equation}
\label{eq::Central_Presusre}
    p_0:=p(0)=\rho_{0}\left[\frac{(1- (1-2GM/R)^{1/2}}{3(1-2GM/R)^{1/2}-1}\right]. 
\end{equation}.} Figure \ref{fig::SUN_Metric_Gamma} shows the results from 
the numerical integration of the above ODEs for the Sun in hydrostatic equilibrium in GR. 
The top left panel depicts the mass function $\mu(r)$ Eq. (\ref{massfunct}). This 
quantity converges to $M_\odot$ just at the Sun's surface. The top right panel
plots the pressure $p(r)$. The bottom left panel shows the metric functions and its product. 
Note that the non-linear treatment together with the accuracy of the numerical implementation allow us to resolve the small deviations of the metric functions $n(r), m(r)$ with respect to the 
Minkowski values which are $\sim 10^{-6}$. Usually this problem, which is associated with the weak field, is treated perturbatively in order to deal only with the small deviations. However, here we solve for the full metric in order to illustrate that our method, both in GR and also in $f(R)$ as discussed in the main text, is able to deal with the weak field limit from the full non-perturbative treatment and prevent us from discarding potentially important nonlinear effects that can take place, specially in $f(R)$ gravity, even in this regime.
On the other hand, and as we argued in the main text, in $R-squared$ gravity the perturbative approaches underestimate the actual values of the NS masses. In that scenario the perturbations are 
with respect to the strong-gravity background in GR. Our treatment can deal with any situation of this kind regardless if the effect of the term $aR^2$ is small or not, and regardless if gravity is weak or strong. This Appendix illustrates this point in the case of GR as opposed with the perturbative approach around the Minkowski background. In this sense, our treatment can deal with weak or strong fields without changing the equations, the numerical code or both.

It is useful to define the PPN $\gamma$  function 
\begin{align}
\label{eq::GAmma_param_GR}
    \gamma &= \Big\vert \frac{1 - m(r)}{1 - n(r)}\Big\vert,
\end{align}
which provides the actual observed value $\gamma \sim 1$ asymptotically.
However, due to the numerical errors in the direct implementation of the above expression in GR 
when dividing  by small numbers \footnote{The metric functions $m$ and $n$ are expected to be of the order $1\pm 2\Phi$, 
respectively, with $\Phi= G M_\odot/{c^2 r}$ the Newtonian potential, and thus, 
$\Phi\sim10^{-6}$ near the Sun's surface, and even lower in the rest of the solar system.}, it is easier to approximate \eqref{eq::GAmma_param_GR} by using $m=n^{-1}$. 
This approximation is exact in GR in vacuum, namely outside the fluid's compact support. Then 
the PPN $\gamma$ becomes\footnote{Strictly speaking, in GR the $\gamma$ parameter is 
defined in the weak field limit and for the Sun's exterior solution, leading to a value that is
exactly one, that is the reason why Eq. \eqref{eq::GAmma_param_GR} is difficult to be evaluated numerically. Nevertheless, the definition \eqref{eq::GAmma_param_GR} is more suitable to measure deviations from GR that are expected in alternatives theories of gravity such as $f(R)$ theories, where {\it a priori} the equality $m=n^{-1}$ is not verified outside the Sun, since the Schwarzschild solution is not necessarily valid when $f(R)\neq R$. For instance, if the chameleon effect does not appear in the solution or if it is 
eliminated by a wrong approximation in the equations, then the exterior solution of an $f(R)$ model 
may lead to $n(r)\approx 1- 2G M_\odot/{c^2 r}$ and $m(r)\approx 1+  G M_\odot/{c^2 r}$ \cite{negrelliSolarSystemTests2020}, which according to \eqref{eq::GAmma_param_GR} yield $\gamma\sim 1/2$ (cf. the analysis of Ref.\cite{chibaSolarSystemConstraints2007} using isotropic coordinates).}  
 \begin{equation}
 \label{eq:GAmma_param_GR_1}
     \gamma = \vert n^{-1}\vert,
 \end{equation}
 so when $n\rightarrow 1$ outside the Sun, $\gamma\rightarrow 1$. The bottom right panel of 
 Fig.\ref{fig::SUN_Metric_Gamma} depicts the deviation, $|1-\gamma|$. 
 We observe that this deviation is already below $\sim 4\times 10^{-4}$ at the Sun's surface, which is the maximum deviation allowed by
the solar system experiments \cite{bertottiTestGeneralRelativity2003}.

\begin{figure}
\centering
\includegraphics[width=\textwidth]{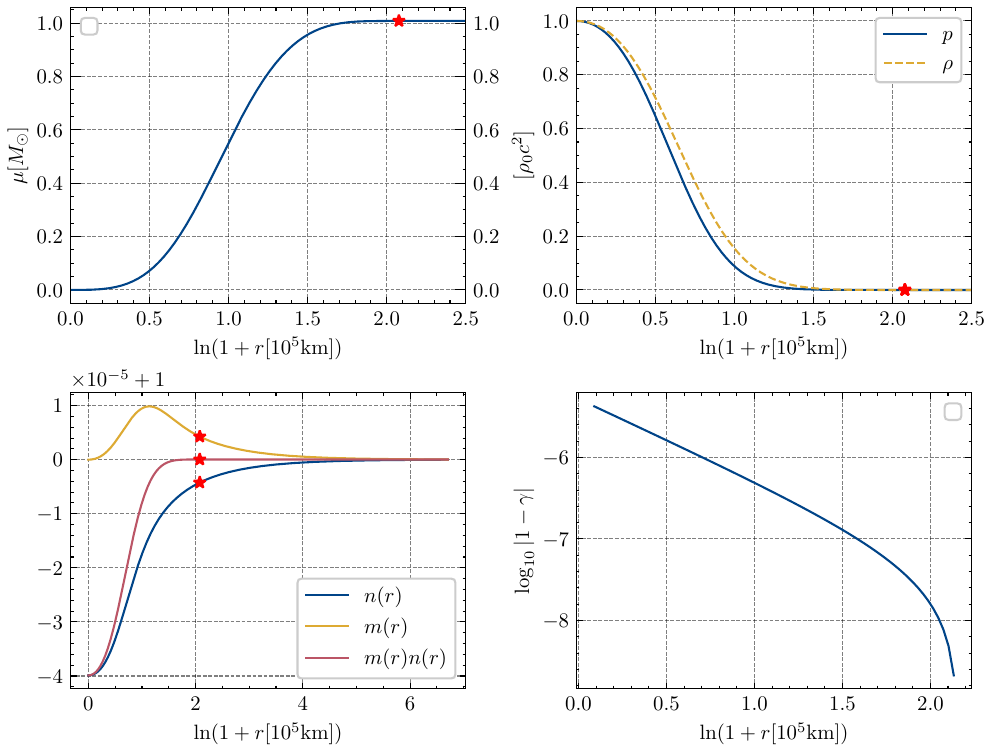}
\caption{
{\it Top left:} Mass function $\mu(r)$ (in units of $M_\odot$) for a simple polytrope with $N=3.35$ ($\Gamma=1+1/N$) that mimics the Sun \cite{hendryPolytropicModelSun1993}. The radial coordinate is in units of  $10^5$ km. The red star indicates the location of the Sun radius. {\it Top right:} Pressure $p(r)$ (solid blue line) and density $\rho c^2$ (yellow dashed line) in units of the 
central value $\rho_0 c^2$, associated with the solution of top-left panel.
{\it Bottom left:} Metric functions  $m(r)$ (solid blue line) and $n(r)$ (solid yellow  line) and 
the product $m(r)\times n(r)$, associated with the solutions of previous 
panels. In comparison to the incompressible fluid, the metric functions are smooth, but their product remains one outside the star indicating a Schwarzschild solution outside the Sun. {\it Bottom right:}  $\gamma$ parameter outside the Sun as defined in 
Eq. \eqref{eq:GAmma_param_GR_1} as a function of the radial coordinate.} 
\label{fig::SUN_POLY_PROFILES}
\end{figure}

As a final note, Fig. \ref{fig::SUN_POLY_PROFILES} displays the solutions of the system \eqref{eq::TOV_m_and_n_variables} when supplemented with a simple polytropic EOS designed to approximate the Sun’s internal structure and reproduce its bulk properties, like the mass and radius. Specifically, we take $p= K \rho ^\Gamma$ with $\Gamma =1.2985$ and $K =p_0/\rho_0^\Gamma $ using the central values $\rho_0=1.53 \times 10^{5} \text{kgm}^{-3}$ and $p_0=3\times10^{16} \text{Nm}^{-2}$ \cite{hendryPolytropicModelSun1993}. Unlike the incompressible fluid for the Sun (cf. Fig. \ref{fig::SUN_Metric_Gamma}), 
where the density is taken as a step function, the polytropic model produces a smoothly varying density throughout the stellar interior (top right panel of Fig. \ref{fig::SUN_POLY_PROFILES}).
 
\section{Piecewise Parametrized Polytropic EOS for NS}
\label{sec::Appendix_Piecewise polytrope EoS}
In this Appendix, we briefly describe the parametrization for the EOS employed in Section \ref{sec:polytrop} and developed by \citet{readConstraintsPhenomenologicallyParameterized2009} to describe NS. This parametrization consists of stitching together polytropic EOS $p=K_i\bar{\rho}^{\Gamma_i}$ ($i=1,2,3\cdots$), 
with different fixed adiabatic index $\Gamma_i$ and $K_i$, over successive density intervals $\bar{\rho}_0 < \bar{\rho}_1<\bar{\rho}_{2}\cdots$, which capture the layering of envelopes and crusts of a NS.\footnote{To keep our notation consistent $\bar{\rho}$ represents the rest mass density, while the quantity $\rho c^2$ which appears in the energy-momentum tensor $T_{ab}= (\rho c^2+ p )u_au_b + g_{ab} p$, is the total energy density. In Ref. \citet{readConstraintsPhenomenologicallyParameterized2009} $\epsilon$, $\rho$ and $e$ corresponds to the  total energy density, rest mass density and internal energy respectively.} Thus, above a dividing density $\bar\rho_0$ we have
\begin{align}
p(\bar{\rho})&=K_i\bar{\rho}^{\Gamma_i}, & d\left( \frac{\rho}{\bar{\rho}}\right)&= - pd\left(\frac{1}{\bar{\rho}}\right), & \bar{\rho}_{i-1}\leq&\bar{\rho}\leq\bar{\rho}_i,
\end{align}
where the second identity comes from the first law of thermodynamics\footnote{For an adiabatic process, i.e., ($dQ=0$) we have $dU = -pdV$ where $U=\rho$V is the total energy of the fluid element and $\rho$ is the energy density. We can rewrite the first law by computing $d\rho$ and substituting back which gives us $d\rho = - \frac{\rho + p}{V}dV$. Introducing the rest mass $M= \bar{\rho} V$ the first law becomes $d\rho = - \frac{\rho + p}{\bar\rho}d\bar{\rho}$ which can be further rewritten in terms of the internal energy per unit mass $\epsilon=\frac{U}{M}=\frac{\rho V}{M}=\frac{\rho}{\bar{\rho}}$ as $d\epsilon = - p d\frac{1}{\bar\rho}$.}. Furthermore, by applying the polytropic relation this relation can be integrated for each density interval as 
\begin{equation}
\label{eq::energydensity_intermsofp}
   c^2\rho^{(i)}:=c^2\rho(\bar\rho)\Big\vert_{\bar{\rho}_{i-1}\leq\bar{\rho}\leq\bar{\rho}_i} = (1 + a_i)\bar{\rho} + \frac{K_i}{\Gamma_i - 1}\bar{\rho}^{\Gamma_i},
\end{equation}
for $\Gamma_i\neq1$. Since we are requiring continuity of both the pressure $p$ and energy density $\rho$, that means at each dividing density $\bar\rho_i$ the following relations holds:
\begin{align}
\rho^{(i)}(\bar\rho_{i})&=\rho^{(i+1)}(\bar\rho_{i}),& p^{(i)}(\bar\rho_{i})&=p^{(i+1)}(\bar\rho_{i}),
\end{align}
which in conjunction with Eq. \eqref{eq::energydensity_intermsofp} and the polytropic relation gives the following recursive relations:
\begin{align}
\label{eq::PPEOS_recursive_relations}
K_{i+1}&=K_i\bar\rho_{i}^{\;\Gamma_i - \Gamma_{i+1}},& a_i&=\frac{\rho^{(i)}(\bar\rho_{i-1})}{\bar\rho_{i-1}} - 1 -\frac{K_i}{\Gamma_i-1}\bar\rho_{i-1}^{\;\Gamma_i -1}.
\end{align}
defining a piecewise parametrized polytropic EOS (PPP EOS). 

In principle, one may construct the EOS from an arbitrary set of polytropes. However, as shown in \cite{readMeasuringNeutronStar2009}, observables such as the mass and radius are relatively insensitive below nuclear saturation density ($\sim 10^{14}\text{g/cm}^3$) regardless of the EOS used. Thus, we can consider two density domains, above and below a matching density  $\bar\rho_0$, chosen of the order of the nuclear saturation density, and match two distinct EOS at  $\bar\rho_0$.

\begin{figure}
\centering
\includegraphics[width=\textwidth]{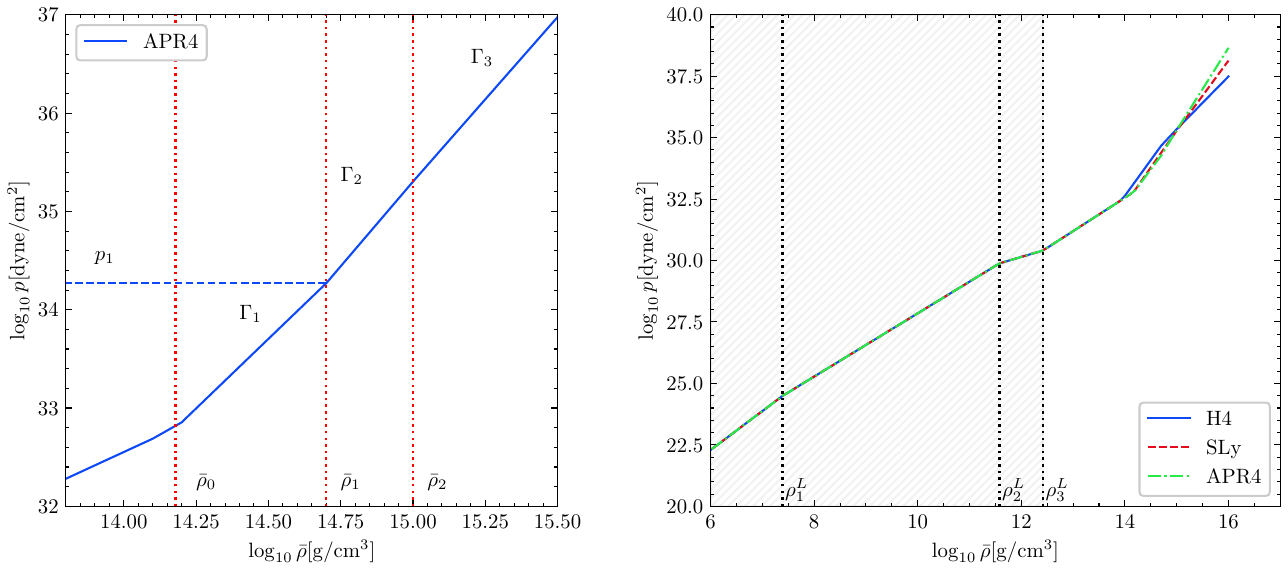}
\caption{\textit{Left panel:} High-density sector for the EOS  APR4 of Table \ref{table::PPP_Models}.  The vertical dotted red lines corresponds to the dividing densities $\bar\rho_1 = 10^{14.7}$ g/cm$^3$ and $\bar\rho_2 = 10^{15.0}$ g/cm$^3$ and $\bar\rho_0 =10^{14.17}$ g/cm$^3$.  \textit{Right panel:} full EOS covering both density domains, high and low, where the low-density part (light shaded region) 
is given by the four polytropic pieces of Table \ref{table::PPP_Models_LOW_DENS}.} 
\label{fig::NGR_PPPpress_Vs_density_PPP_}
\end{figure}

In this work we considered different PPP EOS above the nuclear density. Each one of these choices is matched to a low-density EOS (see paragraph below) by extending the lowest-density polytropic segment until it intersects the low-density EOS (see left panel of Fig. \ref{fig::NGR_PPPpress_Vs_density_PPP_}). Furthermore, each choice of the high-density EOS (above $\bar\rho_0$)
is additionally subdivided in three regions by the two densities $\bar\rho_1 = 10^{14.7}$ g/cm$^3$ and $\bar\rho_2 = 10^{15.0}$ g/cm$^3$. Thus, once the pressure at the first dividing density and the three adiabatic indexes $\{p_1, \Gamma_1, \Gamma_2, \Gamma_3\}$ are specified, the relations \eqref{eq::PPEOS_recursive_relations}, fully determine the high-density PPP EOS. The parameters for the PP models are shown in Table \ref{table::PPP_Models}. 
\begin{table}
\begin{center}
 \begin{tabular}{||c|cccc||} \hline
 EOS & $\log p_1 (\rm{dyne/cm}^{2}$) & $\Gamma_{1}$ & $\Gamma_{2}$ & $\Gamma_{3}$ \\ 
 \hline\hline
APR4 & 34.269 & 2.830 & 3.445 & 3.348 \\ 
SLy &34.348 & 3.005 & 2.988 & 2.851 
\\ 
H4  &34.669 & 2.909 & 2.246 & 2.144  
\\ 
 \hline \hline
\end{tabular}
\end{center}
\caption[table]{Parameters associated with the high-density sector of the APR4, SLy and H4 PPP EOS. These parameters are combined with the values of the low-density part of Table \ref{table::PPP_Models_LOW_DENS} to complete the EOS in both density regions. The data was taken directly from Table III of  Ref. \cite{readConstraintsPhenomenologicallyParameterized2009}. }
\label{table::PPP_Models}
\end{table}

As noted above, the low‐density segment is, in principle, decoupled from the high‐density PPP EOS and may be any cold‐matter model. Here, we follow \citet{readConstraintsPhenomenologicallyParameterized2009} in adopting an analytical representation of the SLy EOS \cite{douchinUnifiedEquationState2001} based on four polytropic pieces between $10^3\text{g/cm}^3 $ and $10^{14}\text{g/cm}^3$. In the same notation described above, Table \ref{table::PPP_Models_LOW_DENS} shows the values $\{\Gamma_i, K_i, \rho_i\}$ for each piece. In conclusion, each of the EOS consists of seven polyropic pieces, of which the first four mimic the lower density SLy, while the last three PPP EOS mimic the high density realistic EOS, the high-density SLy \cite{douchinUnifiedEquationState2001}, APR4 \cite{akmalEquationStateNucleon1998} and H4 \cite{lackeyObservationalConstraintsHyperons2006}. 
As an example of the construction outlined above, Fig.\ref{fig::NGR_PPPpress_Vs_density_PPP_} (left panel) shows the APR4 as mimicked by 
the PPP EOS at high densities.  The right panel depicts the three mimicker PPP EOS at high densities (SLy, H4, APR4) matched with the lower density PPP EOS (shaded region) mimicking the low-density part of SLy.

\begin{table}[!htb]
\begin{center}
\begin{tabular}{||ccc||}
\hline
$K_i$ & $\Gamma_i$ & $\rho_i$ \\
\hline\hline
6.80110e-09 & 1.58425 & 2.44034e+07\\
1.06186e-06 & 1.28733 & 3.78358e+11\\
5.32697e+01 & 0.62223 & 2.62780e+12\\
3.99874e-08 & 1.35692 & --\\
\hline
\end{tabular}
\end{center}
\caption{Parameters for the SLy PPP EOS below nuclear density which is given by four polytropes ($i =1-4$) specified by $\Gamma_i$, $\rho_i$ (in $\mathrm{g/cm^{3}}$) and $K_i$ (in cgs units).
The corresponding value of $p$ is in units of $\mathrm {dyne/cm^2}$. The values are taken from Appendix 1 of \cite{readConstraintsPhenomenologicallyParameterized2009}. 
The last dividing density which does not appear in the table is the density where the low density EOS matches the
high-density EOS and depends on the parameters $p_1$ and $\Gamma_1$ of the high
density EOS for the SLy EOS model (see Fig. \ref{fig::NGR_PPPpress_Vs_density_PPP_}).}
\label{table::PPP_Models_LOW_DENS}
\end{table}

\section{NS solutions in {\it R-squared} gravity with $a<0$}
\label{app:NSanegative}
\begin{figure}
    \centering
    \includegraphics[width=1\linewidth]{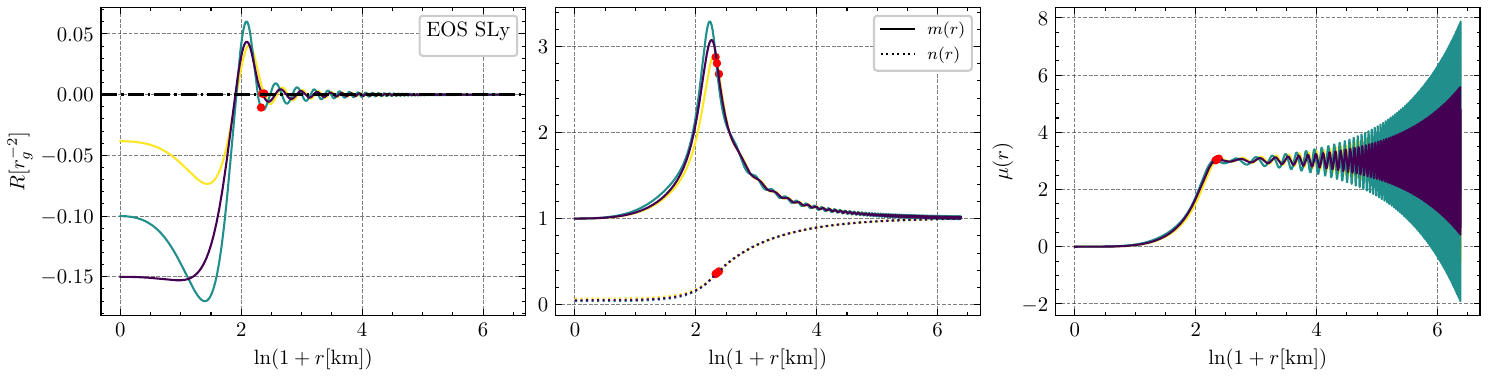}
  \caption{\textit{Left}: Radial profiles of the Ricci scalar for $\bar a=-0.1$ using the SLy EOS. 
The configurations are computed for central densities 
$\rho_0=(2.13,\,2.43,\,2.63)\times10^{15}\ \mathrm{g\,cm^{-3}}$, 
with the central curvature $R_0$ chosen arbitrarily (no shooting is required to achieve this asymptotically fall-off but oscillatory behavior). 
\textit{Middle}: Metric potentials $m(r)$ (solid line) and $n(r)$ (dotted line) corresponding to the same solutions. 
\textit{Right:} Mass function $\mu(r)$ extracted from $m(r)$ as defined by Eq. \eqref{massfunct}. Due to the logarithmic scale in the horizontal axis the oscillation frequency seems to increase with $r$.}
\label{fig:2Negative_a-0.1.pdf}
\end{figure}

In this Appendix we compute NS by taking the {\it R-squared} model (\ref{eq::f_R_RSquered_model}) with $a<0$. Contrary to the claims of Refs.\cite{apariciorescoNeutronStarsTheories2016,fernandezRealisticBuchdahlLimit2025}, their choice $a<0$ is not consistent 
with a possible change of convention defining the Ricci scalar $R$\footnote{Refs.\cite{apariciorescoNeutronStarsTheories2016,fernandezRealisticBuchdahlLimit2025} use different conventions between themselves, for instance, 
Ref.\cite{apariciorescoNeutronStarsTheories2016} use a signature ${\rm (+,-,-,-)}$ and the Lagrangian is written 
in terms of $R + f(R)$, while in Ref.\cite{fernandezRealisticBuchdahlLimit2025} the authors adopt the signature ${\rm (-,+,+,+)}$, like us, and the Lagrangian is also written in terms of $f(R)$, exactly as in the current work.}. For instance, from our Eq. \eqref{eq::fR_formalism_3}, we see that 
in the absence of matter (i.e. in the region outside the NS) the equation reduces to $\Box R - \frac{1}{6a} R = 0$. 
Clearly, by leaving $a$ fixed, this equation remains invariant by any change $R\rightarrow -R$ due to the use of a possible different 
convention in the definition of the Ricci scalar, or the Riemann tensor, for that matter. This invariance is because the equation is 
linear in $R$. Therefore, once the convention 
for $R$ is fixed, this equation is not invariant with respect to the change $a\rightarrow -a$. In our case, taking $a>0$, 
the resulting effective mass $m_{eff}^2=\frac{1}{6a} >0 $. Nevertheless with $a<0$, one has a tachyonic behavior. The latter manifests in the SSS scenario with the following asymptotic behavior $R\sim e^{\pm  \frac{\imath r}{\sqrt{-6a}}}/r$. That is, an 
oscillating $R$ with respect to $r$, regardless of the value $R_0$ at the center of the star. In this case, 
the oscillatory behavior emerges naturally and it cannot be avoided by any shooting method whatsoever.
Notwithstanding, for $a>0$, the asymptotic behavior is $R\sim e^{\pm \frac{r}{\sqrt{6a}}}/r$, and the 
shooting method allows us to eliminate the growing solution and to keep the Yukawa type of solution that decays exponentially 
as $R\sim e^{-\frac{r}{\sqrt{6a}}}/r$ with no oscillatory behavior. In order to proof our assertions, we recover the type of 
solutions found in Refs.\cite{apariciorescoNeutronStarsTheories2016,fernandezRealisticBuchdahlLimit2025} by taking $a<0$ 
in the range $-0.001$ km$^2$ to $-0.05$  km$^2$, and selecting the SLy EOS, although the qualitative results will not change by 
using any other of the two polytropic EOS described in this paper. 
Figure \ref{fig:2Negative_a-0.1.pdf} (left panel) shows a sample of solutions obtained with a given $R_0$ where the explicit oscillatory 
behavior is found as $r\rightarrow \infty$. Note that even if the Ricci scalar falls-off as 
$\sim 1/r$ this is not sufficient for an AF 
spacetime. One would require at least a fall-off $R \sim 1/r^{3+\epsilon}$ ($\epsilon>0$), e.g., $R \sim 1/r^{4}$ to recover the AF behavior. 
As a consequence of this oscillatory behavior, the mass function $\mu (r)$ (right panel)  does not converge to a fixed value asymptotically, 
and therefore, the usual ADM mass is not well defined for this kind of spacetimes.  The oscillatory behavior of $\mu (r)$ manifests also 
in the oscillatory behavior of the metric function $m(r)=g_{rr}$ (or viceversa; middle panel). We have checked that the oscillatory behavior remains for 
other values $a<0$. We conclude that NS solutions with $a<0$ are not 
AF, and that this is not a matter of the conventions used to define $R$, but completely different solutions relative to the more physical 
scenario with $a>0$, where genuinely AF neutron star solutions can be found.

\section{Summary of the different studies found in the analysis of NSs in R-squared gravity}
\label{app:summary}

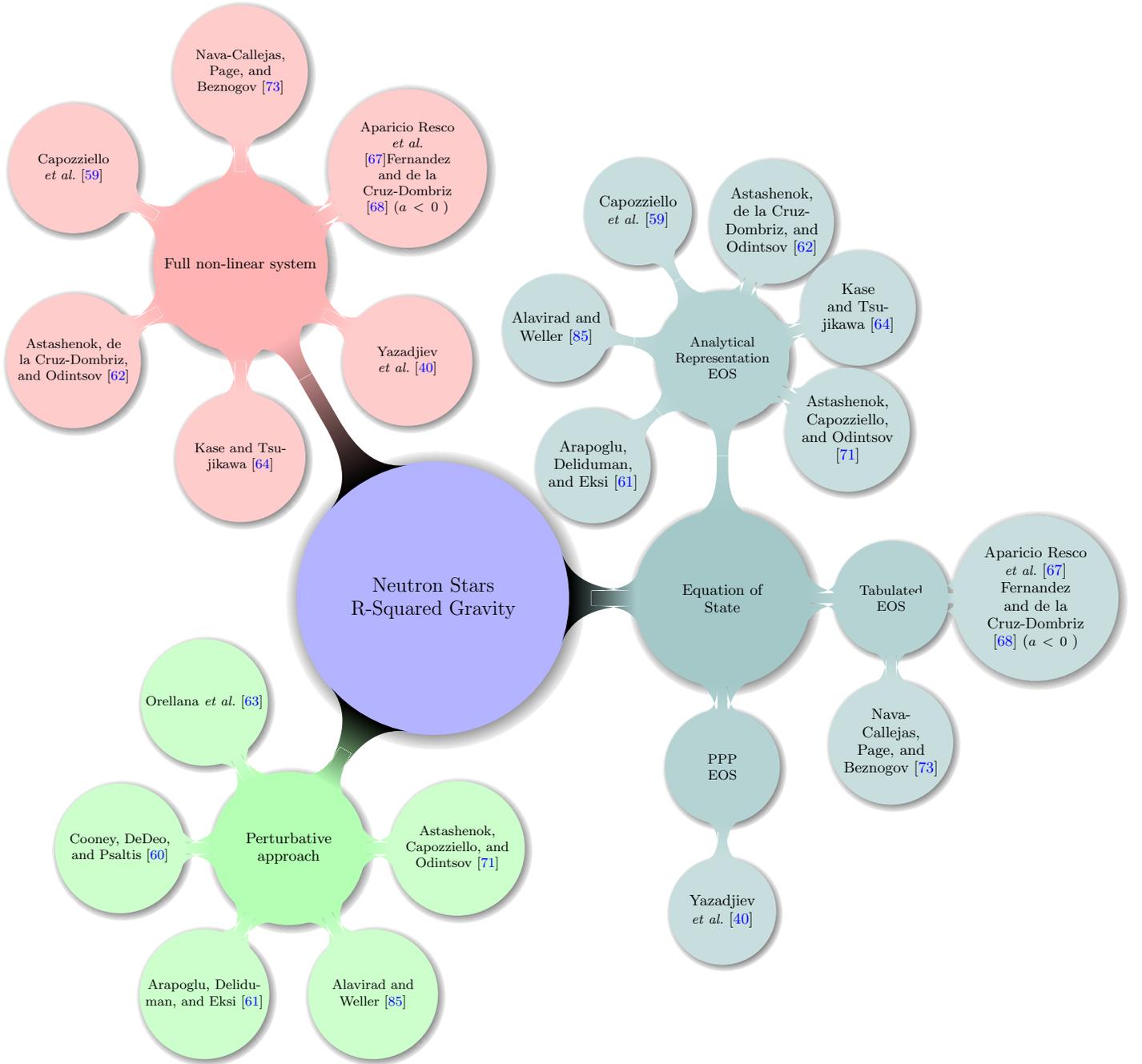
\begin{figure}
 \scalebox{0.8}{ \begin{tikzpicture}[
  mindmap,
  grow cyclic,
  every node/.style={
    concept,
    circular drop shadow,
    minimum size=2.0cm,
    text=black,
    align=center
  },
  every edge/.style={
    draw,
    color=black,
    ->=stealth, 
    shorten >=1pt,
    shorten <=1pt
  },
  level 1 concept/.append style={
    level distance=6cm,
    sibling angle=120
  },
  level 2 concept/.append style={
    level distance=3.5cm
  },
  level 3 concept/.append style={
    font=\fontsize{8pt}{6pt},
    level distance=3cm
  }
]

\begin{scope}[shift={(15cm,0)}]  
\node[concept color=blue!30, text width=5.5cm] (constraints2)
{Neutron Stars\\ R-Squared Gravity}
child[
  concept color=green!30,
  text width=3cm
] {
  node {Perturbative\\ approach}
  child[ text width=2.5cm, concept color=green!20 ]
    { node (perturbSub1) {\citet{orellanaStructureNeutronStars2013}} }
  child[ text width=2.5cm, concept color=green!20 ]
    { node {\citet{cooneyNeutronStarsGravity2010}} }
  child[ text width=2.5cm, concept color=green!20 ]
    { node {\citet{arapogluConstraintsPerturbativeGravity2011}} }
  child[ text width=2.5cm, concept color=green!20 ]
    { node {\citet{alaviradModifiedGravityLogarithmic2013}} }
  child[ text width=2.5cm, concept color=green!20 ]
    { node {\citet{astashenokFurtherStableNeutron2013}} }
}
child[
  concept color=myTeal,
  text width=3.5cm
] {
  node {Equation of\\ State}
  child[ 
    concept color=myTeal!80,
    text width=2.2cm,
    clockwise from=0,  
    level distance=3.5cm
  ] {
    node {Tabulated\\ EOS}
    child[ text width=2.5cm, concept color=myTeal!60, clockwise from=0, level distance=3cm ]
      { node {\citet{apariciorescoNeutronStarsTheories2016} \citet{ fernandezRealisticBuchdahlLimit2025} ($a<0$ )} }
    child[ text width=2.0cm, concept color=myTeal!60, clockwise from=-60, level distance=3cm ]
      { node {\citet{nava-callejasProbingStrongField2023}} }
  }
  child[ 
    concept color=myTeal!80,
    text width=2.5cm,
    clockwise from=150,  
    level distance=5cm
  ] { 
    node {Analytical Representation\\ EOS}
    child[ text width=2.0cm, concept color=myTeal!60, clockwise from=220, level distance=3.5cm ]
      { node {\citet{arapogluConstraintsPerturbativeGravity2011}} }
    child[ text width=2.0cm, concept color=myTeal!60, clockwise from=200, level distance=3.5cm ]
      { node {\citet{alaviradModifiedGravityLogarithmic2013}} }
    child[ text width=2.0cm, concept color=myTeal!60, clockwise from=180, level distance=3.5cm ]
      { node {\citet{capozzielloMassRadiusRelationNeutron2016}} }
    child[ text width=2.0cm, concept color=myTeal!60, clockwise from=160, level distance=3cm ]
      { node {\citet{astashenokRealisticModelsRelativistic2017}} }
    child[ text width=2.0cm, concept color=myTeal!60, clockwise from=140, level distance=3cm ]
      { node {\citet{kaseNeutronStarsGravity2019}} }
    child[ text width=2.0cm, concept color=myTeal!60, clockwise from=120, level distance=3cm ]
      { node {\citet{astashenokFurtherStableNeutron2013}} }
  }
  child[ 
    concept color=myTeal!80,
    text width=2.2cm,
    clockwise from=30,  
    level distance=3.5cm
  ] { 
    node {PPP \\ EOS}
    child[ text width=2.0cm, concept color=myTeal!60, level distance=3cm, clockwise from=-90]
      { node {\citet{yazadjievNonperturbativeSelfconsistentModels2014}} }
  } 
}  
child[
  concept color=red!30,
  text width=3.5cm,
  level distance=8cm
] {
  node {Full non-linear system}
  child[ text width=2.5cm, concept color=red!20, level distance=4cm ]
    { node {\citet{yazadjievNonperturbativeSelfconsistentModels2014}} }
  child[ text width=2.5cm, concept color=red!20, level distance=4cm ]
    { node {\citet{apariciorescoNeutronStarsTheories2016}\citet{ fernandezRealisticBuchdahlLimit2025}  ($a<0$ ) } }
  child[ text width=2.5cm, concept color=red!20, level distance=4cm ]
    { node {\citet{nava-callejasProbingStrongField2023}} }
  child[ text width=2.5cm, concept color=red!20, level distance=4cm ]
    { node {\citet{capozzielloMassRadiusRelationNeutron2016}} }
  child[ text width=2.5cm, concept color=red!20, level distance=4cm ]
    { node {\citet{astashenokRealisticModelsRelativistic2017}} }
  child[ text width=2.5cm, concept color=red!20, level distance=4cm ]
    { node {\citet{kaseNeutronStarsGravity2019}} }
};
\end{scope}


\end{tikzpicture}}
  \caption{Strategies and authors for the analysis of Neutron Stars in $R-squared$ gravity.}
  \label{fig:ns_rsquared}
\end{figure}

The following diagram (Fig. \ref{fig:ns_rsquared})  summarize the main aspects adopted by several authors 
in the analysis of neutron stars in $R-squared$ gravity. The diagram focuses mainly on the approach (perturbative and non-perturbative) and on the different EOS employed 
in those studies. 

\clearpage
\bibliographystyle{aapmrev4-2.bst}
\bibliography{Bibstars}

\end{document}